\def\etc{{\it etc.}}

\def\ie{{\it i.e.}}

\def\~{{$\tilde{\phantom{a}}$}}

\documentclass [12pt] {article}
\usepackage{epsfig}
\usepackage{color}

\newcommand{\postscript}[2]
 {\setlength{\epsfxsize}{#2\hsize}
  \centerline{\epsfbox{#1}}}

\renewcommand{\arraystretch}{1.5}

\def\thebibliography#1{\section{References}\markboth
 {REFERENCES}{REFERENCES}\list
 {[\arabic{enumi}]}{\settowidth\labelwidth{[#1]}\leftmargin\labelwidth
 \advance\leftmargin\labelsep
 \usecounter{enumi}}
 \def\newblock{\hskip .11em plus .33em minus -.07em}
 \sloppy
 \sfcode`\.=1000\relax}
\def\upcite#1{\raise6pt\hbox{\scriptsize
\cite{#1}}}
  \def\lsim{\mathrel {\vcenter {\baselineskip 0pt \kern 0pt
    \hbox{$<$} \kern 0pt \hbox{$\sim$} }}}
    \def\gsim{\mathrel {\vcenter {\baselineskip 0pt \kern 0pt
    \hbox{$>$} \kern 0pt \hbox{$\sim$} }}}

\def\r{\\ [2pt]} 
\def\hline{\noalign{\hrule \vskip2pt}}

\def\|{\ifmmode\Vert\else \char`\|\fi}
\ifx\oldzeta\undefined                          
  \let\oldzeta=\zeta                            
  \def\zzeta{{\raise 2pt\hbox{$\oldzeta$}}}     
  \let\zeta=\zzeta                              
\fi

\ifx\oldchi\undefined                           
  \let\oldchi=\chi                              
  \def\cchi{{\raise 2pt\hbox{$\oldchi$}}}       
  \let\chi=\cchi                                
\fi

\def\frac#1#2{{#1 \over #2}}

\def\half{\ifinner {\scriptstyle {1 \over 2}}
   \else {1 \over 2} \fi}

\def\abs#1{\left\vert#1\right\vert}	

\def\simge{\mathrel{%
   \rlap{\raise 0.511ex \hbox{$>$}}{\lower 0.511ex \hbox{$\sim$}}}}
\def\simle{\mathrel{
   \rlap{\raise 0.511ex \hbox{$<$}}{\lower 0.511ex \hbox{$\sim$}}}}

\def\buildchar#1#2#3{{\null\!                   
   \mathop#1\limits^{#2}_{#3}                   
   \!\null}}                                    
\def\overcirc#1{\buildchar{#1}{\circ}{}}

\def\slashchar#1{\setbox0=\hbox{$#1$}           
   \dimen0=\wd0                                 
   \setbox1=\hbox{/} \dimen1=\wd1               
   \ifdim\dimen0>\dimen1                        
      \rlap{\hbox to \dimen0{\hfil/\hfil}}      
      #1                                        
   \else                                        
      \rlap{\hbox to \dimen1{\hfil$#1$\hfil}}   
      /                                         
   \fi}                                         %

\def\subrightarrow#1{
  \setbox0=\hbox{
    $\displaystyle\mathop{}
    \limits_{#1}$}
  \dimen0=\wd0
  \advance \dimen0 by .5em
  \mathrel{
    \mathop{\hbox to \dimen0{\rightarrowfill}}
       \limits_{#1}}}                           

\def\overlay#1#2{\ifmmode%
\setbox0=\hbox{$#1$}%
\setbox1=\hbox to\wd0{\hss$#2$\hss}\else%
\setbox0=\hbox{#1}%
\setbox1=\hbox to\wd0{\hss#2\hss}\fi%
#1\hskip-\wd0\box1 }

\def\pmb#1{\leavevmode\setbox0=\hbox{#1}%
\kern-.02em\copy0\kern-\wd0
\kern.04em\copy0\kern-\wd0
\kern-.02em\raise.04em\box0 }

\def\vereq#1#2{\lower3pt\vbox{\baselineskip1.5pt \lineskip1.5pt
\ialign{$\m@th#1\hfill##\hfil$\crcr#2\crcr\sim\crcr}}}

\def\tensor#1{\protect\@ontopof{#1}{\leftrightarrow}{1.15}\mathord{\box2}}
\def\overstar#1{\protect\@ontopof{#1}{\ast}{1.15}\mathord{\box2}}
\def\overdots#1{\protect\@ontopof{#1}{\cdots}{1.0}\mathord{\box2}}
\def\overcirc#1{\protect\@ontopof{#1}{\circ}{1.2}\mathord{\box2}}
\def\loarrow#1{\protect\@ontopof{#1}{\leftarrow}{1.15}\mathord{\box2}}
\def\roarrow#1{\protect\@ontopof{#1}{\rightarrow}{1.15}\mathord{\box2}}

\def\@ontopof#1#2#3{%
{\mathchoice
{\@@ontopof{#1}{#2}{#3}\displaystyle\scriptstyle}%
{\@@ontopof{#1}{#2}{#3}\textstyle\scriptstyle}%
{\@@ontopof{#1}{#2}{#3}\scriptstyle\scriptscriptstyle}%
{\@@ontopof{#1}{#2}{#3}\scriptscriptstyle\scriptscriptstyle}%
}%
}

\def\@@ontopof#1#2#3#4#5{%
\setbox0=\hbox{$#4#1$}%
\setbox1=\hbox{$#5#2$}%
\setbox2=\hbox{}\ht2=\ht0 \dp2=\dp0 %
\ifdim\wd0>\wd1 %
\setbox1=\hbox to\wd0{\hss\box1\hss}%
\mathord{\rlap{\raise#3\ht0\box1}\box0}%
\else   %
\setbox1=\hbox to.9\wd1{\hss\box1\hss}%
\setbox0=\hbox to\wd1{\hss$#4\relax#1$\hss}%
\mathord{\rlap{\copy0}\raise#3\ht0\box1}%
\fi
}%

\def\lambdabar{\protect\@lambdabar}
\def\@lambdabar{%
\relax
\bgroup
\def\@tempa{\hbox{\raise.73\ht0
\hbox to0pt{\kern.25\wd0\vrule width.5\wd0
height.1pt depth.1pt\hss}\box0}}%
\mathchoice{\setbox0\hbox{$\displaystyle\lambda$}\@tempa}%
{\setbox0\hbox{$\textstyle\lambda$}\@tempa}%
{\setbox0\hbox{$\scriptstyle\lambda$}\@tempa}%
{\setbox0\hbox{$\scriptscriptstyle\lambda$}\@tempa}%
\egroup
}

\def\corresponds{{\lower.2ex\hbox{=}}{\rm\kern-.75em^\triangle}}
\def\succsim{\succ\kern-.9em_\sim\kern.3em}
\def\precsim{\prec\kern-1em_\sim\kern.3em}
\def\slantfrac#1#2{\kern1em^{#1}\kern-.3em/\kern-.1em_{#2}}

\begin{document}                                                                

\baselineskip=15pt

\renewcommand{\arraystretch}{1.5}

\begin{center}
{\Large\bf Circular Orbits Inside the Sphere of Death}

\bigskip

{Kirk T.~McDonald}

\medskip

{\sl Joseph Henry Laboratories, Princeton University, Princeton, New Jersey 08544}

\bigskip

(November 8, 1993) 


\bigskip

{\large\bf Abstract}
\end{center}

A wheel or sphere rolling without slipping on the inside of a sphere in a
uniform gravitational field can have stable
circular orbits that lie wholly above
the ``equator", while a particle sliding freely cannot.

\bigskip

\section{Introduction}

In a recent article \cite{Wall} in this Journal, 
Abramowicz and Szuszkiewicz remarked on
an interesting analogy between orbits above the equator of a ``wall of death''
and orbits near a black hole; namely that the centrifugal force in both
cases appears to point towards rather than away from the center of an
appropriate coordinate system.  Here we take a ``wall of death" to be a hollow
sphere on the Earth's surface large enough that a motorcycle can be driven
on the inside of the sphere.  The intriguing question is whether there
exist stable orbits for the motorcycle that lie entirely above the
equator (horizontal great circle) of the sphere.

In ref.~\cite{Wall} the authors stated that no such orbits are possible, perhaps
recalling the well-known result for a particle sliding freely on the
inside of a sphere in a uniform gravitational field.  However, the
extra degrees of freedom associated with a rolling wheel (or sphere)
actually do permit such orbits, in apparent defiance of intuition.
In particular, the friction associated with the condition of rolling without
slipping can in some circumstances have an upward component large enough to
balance all other downward forces.

In this paper we examine the character of all circular orbits inside a
fixed sphere, for both wheels and spheres that roll without slipping.
The rolling constraint is velocity dependent (non-holonomous), so explicit use 
of  a Lagrangian is not especially effective.  Instead we follow a vectorial
approach as advocated by Milne (Chap.~17) \cite{Milne}. This approach does 
utilize the rolling constraint, a
careful choice of coordinates, and the elimination of the constraint force
from the equations of motion, all of which are
implicit in Lagrange's method.  The vector approach is, of course, a convenient
codification of earlier methods in which individual components were
explicitly written out.  Compare with classic works such as those of 
Lamb (Chap.~9) \cite{Lamb}, Deimel (Chap.~7) \cite{Deimel}
and Routh (Chap.~5) \cite{Routh}.

Once the solutions are obtained in sec.~2 for rolling wheels
we make a numerical evaluation of the magnitude of the acceleration in $g$'s,
and of the required coefficient of static friction on some representative
orbits.  The resulting parameters are rather extreme, and the circus name
``sphere of death'' seems apt.

The stability of steady orbits of wheels is considered in some detail, but
completely general results are not obtained (because the general motion has
four degrees of freedom).  All vertical orbits are shown to be stable, as are
horizontal orbits around the equator of the sphere.
We also find that all horizontal orbits away from the poles are stable
in the limit of small wheels, and conjecture that the a similar condition
holds for ``death-defying'' orbits of large wheels above the equator of the 
sphere. In sec.~4 we lend support to this conjecture by comparing to the 
related case of a sphere rolling within a sphere for which a complete
stability analysis can be given.

Discussions of wheels and spheres rolling outside a fixed sphere are given
in secs.~3 and 5, respectively.

\section{Wheel Rolling Inside a Fixed Sphere}

\subsection{Generalities}

\begin{figure}[htp]  
\postscript{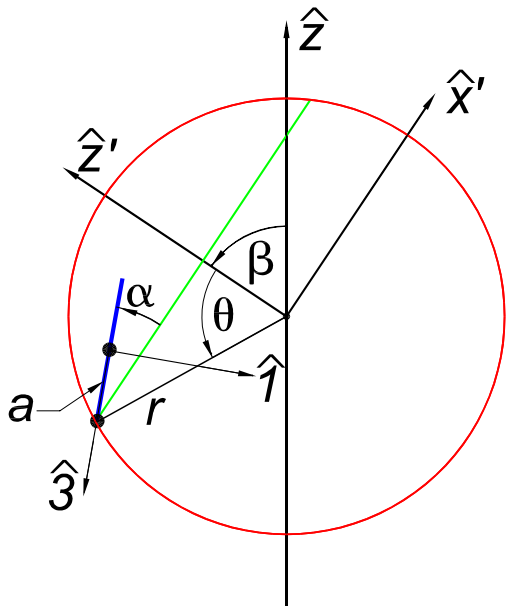}{0.4}
\begin{center}
\parbox{5.5in} 
{\caption[ Short caption for table of contents ]
{\label{fig1} A wheel of radius $a$ rolls without slipping on a circular orbit
inside a fixed sphere of radius $r$.  The orbit sweeps out a cone of angle 
$\theta$
about the $z'$-axis, which axis makes angle $\beta$ to the vertical.  The
$x'$-axis is orthogonal to the $z'$-axis in the $z$-$z'$ plane
The angle between the plane of the orbit and diameter of the wheel
that includes the point of contact with the sphere is denoted by $\alpha$.  
A right-handed
triad of unit vectors, $(\hat{\bf 1},\hat{\bf 2},\hat{\bf 3})$, is defined with
$\hat{\bf 1}$ along the axis of the wheel and $\hat{\bf 3}$ pointing from the
center of the wheel to the point of contact.
}}
\end{center}
\end{figure}

We consider a wheel of radius $a$ rolling without slipping
on a circular orbit on the inner surface of a sphere of radius $r > a$.
The analysis is performed in the lab frame, in which the sphere is fixed.
The $z$-axis is vertical and upwards with origin at the center of the sphere
as shown in Fig.~\ref{fig1}.
As the wheel rolls on the sphere, the point of contact traces a path that is
an arc of a circle during any short interval.  In steady motion the path forms
closed circular orbits which are of primary interest here.  We therefore
introduce a set of axes $(x',y',z')$ that are related to the circular motion
of the point of contact.  If the motion is steady, these axes are fixed in
the lab frame.

\begin{figure}[htp]  
\postscript{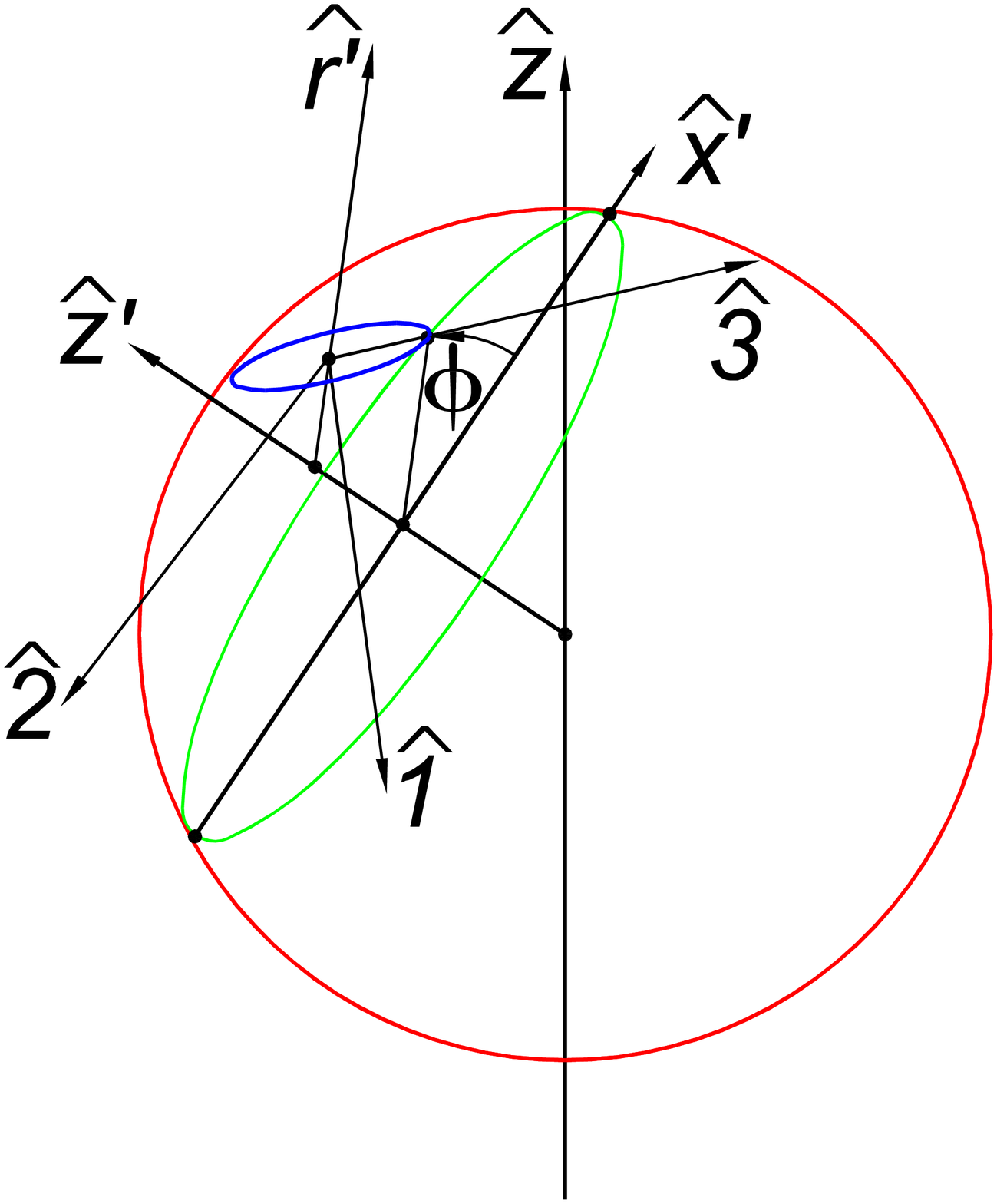}{0.4}
\begin{center}
\parbox{5.5in} 
{\caption[ Short caption for table of contents ]
{\label{fig2} The azimuth of the point of contact of the wheel with the
sphere to the $x'$-axis is $\phi$.  
The unit vector $\hat{\bf r}'$ is orthogonal to the $z'$-axis
and points towards the center of the wheel  (or equivalently, towards the
point of contact).
Unit vector $\hat{\bf 2} = \hat{\bf 3} \times \hat{\bf 1} = \hat{\bf z}'
\times \hat{\bf r}'$.
}}
\end{center}
\end{figure}

The 
normal to the plane of the circular orbit through the center of the sphere (and 
also through the center of the circle) is labeled $z'$.  The angle between
axes $z$ and $z'$ is $\beta$ with $0 \leq \beta \leq \pi/2$.  
A radius from the center of the sphere to the point of contact of the wheel
sweeps out a cone of angle $\theta$ about the $z'$ axis, where $0 \leq
\theta \leq \pi$.  The azimuthal
angle of the point of contact on this cone is called $\phi$, with $\phi = 0$
defined by the direction of the $x'$-axis, which is along the projection of the 
$z$-axis onto the plane of the orbit, as shown in Fig.~\ref{fig2}.
  Unit vectors are labeled with a superscript $\hat{\phantom{a}}$, 
so that $\hat{\bf y}' = \hat{\bf z}' \times \hat{\bf x}'$ completes
the definition of the $'$-coordinate system.

For a particle sliding freely, the only stationary orbits have $\beta = 0$
(horizontal circles) or $\beta = \pi/2$ (vertical great circles). 
For wheels and spheres rolling inside a sphere it turns out that $\beta = 0$
or $\pi/2$ also, as we will demonstrate.  However, the friction at the
point of contact in the rolling cases permits orbits with a larger range of
$\theta$ than in the sliding case.
If $\beta = 0$ or $\pi/2$ were accepted as an
assumption the derivation could be shortened somewhat. 

We also introduce a right-handed 
coordinate triad of unit vectors $(\hat{\bf 1},\hat{\bf 2},\hat{\bf 3})$ related
to the geometry of the wheel.  Axis $\hat{\bf 1}$ lies along the
symmetry axis of the wheel as shown shown in Fig.~\ref{fig1}.
Axis $\hat{\bf 3}$ is directed from the center of the wheel to the point of
contact of the wheel with the sphere.  
The vector from the center of the wheel to the point of contact is then
\begin{equation}
{\bf a} = a \hat{\bf 3}.
\label{eq00}
\end{equation}
Axis $\hat{\bf 2} = \hat{\bf 3} \times \hat{\bf 1}$
lies in the plane of the wheel, and also in the plane of the orbit (the
$x'$-$y'$ plane).  The sense of axis $\hat{\bf 1}$ is chosen  so that the
 component
$\omega_1$ of the angular velocity vector $\vec \omega$  of the wheel about 
this axis is positive.  
Consequently, axis $\hat{\bf 2}$ points in the direction of the velocity of
the point of contact, and therefore is parallel to the tangent to the orbit.

Except for axis $\hat{\bf 1}$, these rotating axes are not body axes,
but the inertia tensor is diagonal with respect to them.  We write
\begin{equation}
I_{11} = 2kma^2, \qquad I_{22} = kma^2 = I_{33},
\label{eq1}
\end{equation}
which holds for any circularly symmetric disc according to the
perpendicular axis theorem; $k = 1/2$ for a wheel of radius $a$ with
mass $m$ concentrated at the rim, $k = 1/4$ for a uniform disc, \etc\

The wheel does not necessarily lie in the plane of the orbit.  Indeed, it
is the freedom to ``bank'' the wheel that makes the ``death-defying''
orbits possible.  The diameter of the wheel through the point of contact
(\ie, axis $\hat{\bf 3}$)
makes angle $\alpha$ to the plane of the orbit.  In general, a wheel can
have an arbitrary rotation about the $\hat{\bf 3}$-axis, but the wheel will
roll steadily along a closed circular orbit orbit only if angular velocity
component $\omega_3$ is such that the plane of
the wheel intersects the plane of the orbit along the tangent to the orbit
at the point of contact.
Hence, for steady motion we will be able to deduce a constraint on $\omega_3$.
The case of a rolling sphere is distinguished by
the absence of this constraint, as considered later.

\begin{figure}[htp]  
\postscript{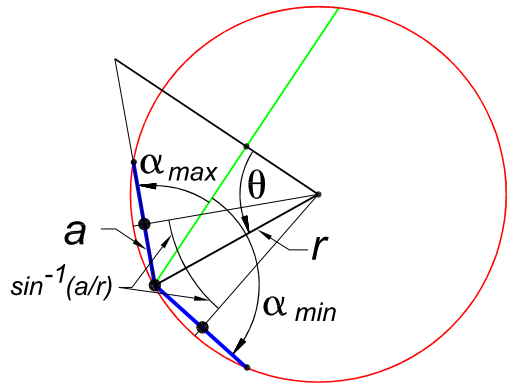}{0.4}
\begin{center}
\parbox{5.5in} 
{\caption[ Short caption for table of contents ]
{\label{fig3} Geometry illustrating the extremes of angle $\alpha$.
}}
\end{center}
\end{figure}

Since the wheel lies
inside the sphere, as shown in Fig.~\ref{fig3}, we can readily deduce the 
geometric relation that
\begin{equation}
\theta - \pi + \sin^{-1}(a/r) < \alpha < 
\theta - \sin^{-1}(a/r).
\label{eq0}
\end{equation}

\begin{figure}[htp]  
\postscript{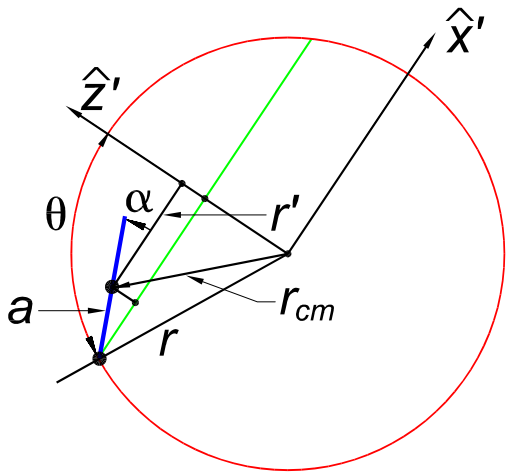}{0.4}
\begin{center}
\parbox{5.5in} 
{\caption[ Short caption for table of contents ]
{\label{fig4} Geometry illustrating the vector ${\bf r}_{\rm cm}$ from the 
center of the sphere to the center of the wheel, and the distance 
$r' = r\sin\theta - a\cos\alpha$ from the $z'$-axis to the center of the wheel.
}}
\end{center}
\end{figure}

It is useful to introduce ${\bf r}' = r' \hat{\bf r}'$ as the
perpendicular vector from the $z'$-axis to the center of the wheel.  
The magnitude $r'$ is given by
\begin{equation}
r' = r\sin\theta - a\cos\alpha,
\label{eq8b}
\end{equation}
as shown in Fig.~\ref{fig4}.  
The vector $\hat{\bf z}' \times \hat{\bf r}'$ is in the direction
of motion of the point of contact, which was defined previously to be
direction $\hat{\bf 2}$.  That is, $(\hat{\bf r}',\hat{\bf 2},\hat{\bf z}')$ form
a right-handed unit triad, which is related to the triad
$(\hat{\bf 1},\hat{\bf 2},\hat{\bf 3})$ by
\begin{equation}
\hat{\bf z}' = -\cos\alpha \hat{\bf 1} - \sin\alpha \hat{\bf 3},
\label{eq11}
\end{equation}
and 
\begin{equation}
\hat{\bf r}' = \hat{\bf 2} \times \hat{\bf z}'
= -\sin\alpha \hat{\bf 1} + \cos\alpha \hat{\bf 3},
\label{eq11a}
\end{equation}
as can be seen from Fig.~1.

The length $r'$ is negative when the center of
the wheel is on the opposite side of the $z'$-axis from the point of contact.
This can occur for large enough $a/r$ when the point of contact is near the
$z'$-axis, such as when $\theta \approx 0$ and $\alpha < 0$ or $\theta \approx
\pi$ and $\alpha > 0$.

The force of contact of the sphere on the wheel is
labeled ${\bf F}$.  For the wheel to be in contact with the sphere the force
${\bf F}$ must have a component towards the center of the sphere,
which will be verified after the motion is obtained.

The equation of motion of the center of mass of the wheel is
\begin{equation}
m {d^2 {\bf r}_{\rm cm} \over dt^2} = {\bf F} - mg\hat{\bf z},
\label{eq2}
\end{equation}
where $g$ is the acceleration due to gravity.
The equation of motion for the angular momentum ${\bf L}_{\rm cm}$ about the center
of mass is
\begin{equation}
{d{\bf L}_{\rm cm} \over dt} = {\bf N}_{\rm cm} = {\bf a} \times {\bf F}.
\label{eq3}
\end{equation}
We eliminate the unknown force ${\bf F}$ in eq.~(\ref{eq3}) via 
eqs.~(\ref{eq00}) and (\ref{eq2}) to find
\begin{equation}
{1 \over ma} {d{\bf L}_{\rm cm} \over dt} = g \hat{\bf 3} \times \hat{\bf z} + 
\hat{\bf 3} \times {d^2 {\bf r}_{\rm cm} \over dt^2}.
\label{eq4}
\end{equation}

The constraint that the wheel rolls without slipping relates the velocity of
the center of mass to the angular velocity vector $\vec\omega$ of the wheel.
In particular, the velocity vanishes for that point on the wheel instantaneously
in contact with the sphere:
\begin{equation}
{\bf v}_{\rm contact} = {\bf v}_{\rm cm} + \vec\omega \times {\bf a} = 0,
\label{eq5}
\end{equation}
and hence
\begin{equation}
{\bf v}_{\rm cm} = {d{\bf r}_{\rm cm} \over dt} = a \hat{\bf 3} \times \vec\omega.
\label{eq6}
\end{equation}
Multiplying this equation by $\hat{\bf 3}$, we find
\begin{equation}
\vec\omega = - \hat{\bf 3} \times {{\bf v}_{\rm cm} \over a} 
+ \omega_3 \hat{\bf 3}.
\label{eq6a}
\end{equation}
Equations (\ref{eq2})-(\ref{eq6a}) 
hold whether the rolling object is a wheel or a sphere.

The strategy now is to extract as much information as possible about the
angular velocity $\vec \omega$ before confronting the full equation of
motion (\ref{eq4}).  
The angular velocity can also be written in terms of the unit vector 
$\hat{\bf 1}$ along the symmetry axis of the wheel as
\begin{equation}
\vec \omega = \omega_1 \hat{\bf 1} + \hat{\bf 1} \times {d\hat{\bf 1} \over dt}.
\label{eq7}
\end{equation}
This follows on writing $\vec\omega = \omega_1 \hat{\bf 1} + \vec\omega_\perp$,
and noting that the rate of change of the body 
vector $\hat{\bf 1}$ 
is just $d\hat{\bf 1}/dt = \vec\omega_\perp \times \hat{\bf 1}$, 
so $\vec\omega_\perp
= \hat{\bf 1} \times d\hat{\bf 1}/dt$.  Using 
eq.~(\ref{eq1}), the angular momentum can now be written as
\begin{equation}
{\bf L}_{\rm cm} = \vec{\vec I} \cdot \vec \omega = 2kma^2 \omega_1 \hat{\bf 1} 
+ kma^2 \hat{\bf 1} \times {d\hat{\bf 1} \over dt}.
\label{eq7b}
\end{equation}

\subsection{Steady Motion in a Circle}

To obtain additional relations we restrict our attention to
orbits in which the point of contact of the wheel with the sphere moves in a
closed circle.  In such cases the center of
mass of the wheel (and also the coordinate triad 
($\hat{\bf 1},\hat{\bf 2},\hat{\bf 3}$)) has
angular velocity $\dot\phi$ about the $\hat{\bf z}'$-axis (and no other component), 
where the dot means differentiation with respect to time.  Thus
\begin{equation}
{\bf v}_{\rm cm} = \dot\phi \hat{\bf z}' \times r' \hat{\bf r}' 
= r'\dot\phi \hat{\bf 2}.
\label{eq8}
\end{equation}

Equation (\ref{eq6a}) can now be evaluated, yielding
\begin{equation}
\hat\omega = (r'/a)\dot\phi \hat{\bf 1}  + \omega_3 \hat{\bf 3}.
\label{eq9}
\end{equation}
For steady motion there can be no rotation about axis $\hat{\bf 2}$; angle
$\alpha$ is constant. To find $\omega_3$ we now pursue eq.~(\ref{eq7}).

As argued above, the angular velocity $\vec\gamma$ of the triad 
$(\hat{\bf 1},\hat{\bf 2},\hat{\bf 3})$ is 
\begin{equation}
\vec\gamma = \dot\phi \hat{\bf z}' 
= -\dot\phi \cos\alpha \hat{\bf 1} - \dot\phi \sin\alpha \hat{\bf 3},
\label{eq7a}
\end{equation}
using eq.~(\ref{eq11}).  Then, 
\begin{equation}
{d\hat{\bf 1} \over dt} = \vec\gamma \times \hat{\bf 1} 
= -\dot\phi \sin\alpha \hat{\bf 2} 
\label{eq10}
\end{equation}
\begin{equation}
{d\hat{\bf 2} \over dt} = \vec\gamma \times \hat{\bf 2}
= \dot\phi \sin\alpha \hat{\bf 1} - \dot\phi \cos\alpha \hat{\bf 3}
= - \dot\phi \hat{\bf r}',
\label{eq17a}
\end{equation}
and
\begin{equation}
{d\hat{\bf 3} \over dt} = \vec\gamma \times \hat{\bf 3} 
= \dot\phi \cos\alpha \hat{\bf 2}.
\label{eq16}
\end{equation}
It immediately follows that
\begin{equation}
\hat{\bf 1} \times {d\hat{\bf 1} \over dt} = - \dot\phi \sin\alpha \hat{\bf 3}.
\label{eq12}
\end{equation}
Comparing with eq~(\ref{eq7}) we see that $\omega_3 = -\dot\phi \sin\alpha$
and hence from eq.~(\ref{eq9}) we find
\begin{equation}
\vec\omega = (r'/a)\dot\phi \hat{\bf 1} - \dot\phi \sin\alpha \hat{\bf 3}.
\label{eq13}
\end{equation}
As anticipated, the rolling constraint 
specifies how $\omega_1$ and $\omega_3$ are both related to
the angular velocity $\dot\phi$ of the wheel about the $\hat{\bf z}'$-axis.

For use in the equation of motion (\ref{eq4}) we can now write
\begin{equation}
{\bf L} = \vec{\vec I} \cdot \vec \omega 
= kma^2[ 2(r'/a)\dot\phi \hat{\bf 1} - \dot\phi \sin\alpha \hat{\bf 3}],
\label{eq14}
\end{equation}
and hence,
\begin{equation}
{1 \over ma}{d{\bf L} \over dt} = 2kr' \ddot\phi \hat{\bf 1} 
  - k\dot\phi^2\sin\alpha (2r' + a \cos\alpha) \hat{\bf 2}
  - ka\ddot\phi \sin\alpha \hat{\bf 3},
\label{eq15}
\end{equation}
using eqs.~(\ref{eq10}-\ref{eq16}).
Also, by differentiating eq. (\ref{eq8}) we find
\begin{equation}
{d^2 {\bf r}_{\rm cm} \over dt^2} = 
r'\dot\phi^2 \sin\alpha \hat{\bf 1} + r' \ddot\phi \hat{\bf 2} 
            - r'\dot\phi^2 \cos\alpha \hat{\bf 3},
\label{eq17}
\end{equation}
so that
\begin{equation}
\hat{\bf 3} \times {d^2 {\bf r}_{\rm cm} \over dt^2} = 
- r' \ddot\phi \hat{\bf 1} + r' \dot\phi^2 \sin\alpha \hat{\bf 2}.
\label{eq18}
\end{equation}
Combining (\ref{eq4}), (\ref{eq15}) and (\ref{eq18}),
 the equation of motion reads
\begin{eqnarray}
g \hat{\bf z} \times \hat{\bf 3} & = & \hat{\bf 3} \times 
{d^2 {\bf r}_{\rm cm}\over dt^2} -
{1 \over ma}{d {\bf L} \over dt} \nonumber \\
& = & -(2k + 1) r' \ddot\phi \hat{\bf 1} 
  + [(2k + 1)r' + k a \cos\alpha] \dot\phi^2 \sin\alpha \hat{\bf 2} 
  + ka\ddot\phi \sin\alpha \hat{\bf 3}.  
\label{eq19}
\end{eqnarray}

To evaluate $\hat{\bf z} \times \hat{\bf 3}$, we first express $\hat z$ in terms 
of the triad ($\hat{\bf r}',\hat{\bf 2},\hat{\bf z}'$), and then transform to 
triad ($\hat{\bf 1},\hat{\bf 2},\hat{\bf 3}$).  
When the point of contact of the wheel (and hence the $\hat{\bf r}'$-axis)
 has azimuth $\phi$ relative to
the $\hat{\bf x}'$ axis, the $\hat{\bf z}$ axis has azimuth $-\phi$
relative to the $\hat{\bf r}'$ axis.  
Hence,
\begin{eqnarray}
\hat{\bf z} & = & \sin\beta\cos\phi \hat{\bf r}' - \sin\beta\sin\phi \hat{\bf 2}
 + \cos\beta \hat{\bf z}' \\
& = & -(\cos\alpha\cos\beta + \sin\alpha\sin\beta\cos\phi) \hat{\bf 1} 
      - \sin\beta\sin\phi \hat{\bf 2} 
      - (\sin\alpha\cos\beta - \cos\alpha\sin\beta\cos\phi) \hat{\bf 3},
\nonumber 
\label{eq21} 
\end{eqnarray}
using eqs.~(\ref{eq11})-(\ref{eq11a}).  Thus,
\begin{equation}
\hat{\bf z} \times \hat{\bf 3} =  - \sin\beta\sin\phi \hat{\bf 1} +
    (\cos\alpha\cos\beta + \sin\alpha\sin\beta\cos\phi) \hat{\bf 2}.
\label{eq22} 
\end{equation} 

The $\hat{\bf 1}$, $\hat{\bf 2}$ and $\hat{\bf 3}$ components of the equation of
motion are now
\begin{equation}
(2k + 1) r' \ddot\phi = g\sin\beta\sin\phi,
\label{eq23}
\end{equation}
\begin{equation}
[(2k + 1)r' + k a \cos\alpha] \dot\phi^2 \sin\alpha 
        = g(\cos\alpha\cos\beta + \sin\alpha\sin\beta\cos\phi),
\label{eq24}
\end{equation}
and
\begin{equation}
ka \ddot\phi \sin\alpha = 0.
\label{eq25}
\end{equation}
The cone angle $\theta$ enters the equations of motion only through $r'$.

\subsubsection{Vertical Orbits}

From eq.~(\ref{eq25}) we learn that for circular orbits either $\sin\alpha = 0$ 
or $\ddot\phi = 0$.  We first consider the simpler case that $\sin\alpha = 0$,
which implies that the plane of the wheel lies in the plane of the orbit.
For a wheel inside the sphere with $\sin\alpha
= 0$, we must have $\alpha = 0$ to satisfy the geometric constraint (\ref{eq0}).
Then eq.~(\ref{eq24}) can only be satisfied if $\cos\beta = 0$; \ie, $\beta =
\pi/2$ and the plane of the orbit is vertical.  The remaining equation of
motion (\ref{eq23}) now reads
\begin{equation}
(2k + 1)r' \ddot\phi = g\sin\phi,
\label{eq26}
\end{equation}
with $r' = r\sin\theta - a > 0$, which integrates to
\begin{equation}
{2k + 1 \over 2} mr'^2 (\dot\phi^2 - \dot\phi_0^2) = mgr' (1 - \cos\phi),
\label{eq27}
\end{equation}
where $\dot\phi_0$ is the angular velocity at the top of the orbit at which
$\phi = 0$.  Equation (\ref{eq27}) expresses conservation of energy.
The angular velocity $\vec\omega$ and
the angular momentum ${\bf L}_{\rm cm}$ vary in magnitude but are always
perpendicular to the plane of the orbit.

The requirement that the wheel stay in contact with the sphere is that
the contact force ${\bf F}$ have component $F_\perp$ that points to the center
of the sphere.
On combining eqs.~(\ref{eq2}), 
(\ref{eq17}), (\ref{eq21}) and (\ref{eq26}) we find
\begin{equation}
{\bf F} = {2k \over 2k + 1} mg \sin\phi \hat{\bf 2} + 
         m(g\cos\phi - r'\dot\phi^2) \hat{\bf 3}.
\label{eq28}
\end{equation}
The contact force is in the plane of the orbit, so the resulting torque about
the center of mass of the wheel changes the magnitude but not the direction
of the angular momentum.
On the vertical orbits, axis $\hat{\bf 2}$ is tangent to the sphere, and axis
$\hat{\bf 3}$ makes angle $\pi/2 - \theta$ to the radius from the center of the
sphere to the point of contact.  
Hence 
\begin{equation}
F_\perp = -F_3\sin\theta
\label{eq28d}
\end{equation}
 is positive
and the orbit is physical so long as the angular velocity $\dot\phi_0$ at the 
peak of the orbit obeys 
\begin{equation}
\dot\phi^2_0 > {g \over r'},
\label{eq28a}
\end{equation}
as readily deduced from elementary considerations as well.

The required
coefficient $\mu$ of static friction is given by $\mu = F_\parallel/F_\perp$
where 
\begin{equation}
F_\parallel = \sqrt{F_3^2\cos\theta^2 + F_2^2}
\label{eq28c}
\end{equation}
is the component of the contact force parallel to
the surface of the sphere.  We see that 
\begin{equation}
\mu = \cot\theta \sqrt{1 + (F_2/F_3\cos\theta)^2},
\label{eq28b}
\end{equation}
which must be greater than $\cot\theta$,
but only much greater if the wheel nearly loses contact at the top of the
orbit.  Hence orbits with $\pi/4 \lsim \theta \leq \pi/2$ are consistent
with the friction of typical rubber wheels, namely $\mu \lsim 1$.

Because a wheel experiences friction at the point of contact, vertical
orbits are possible with $\theta < \pi/2$. This is in contrast to the case of a
particle sliding freely on the inside of a sphere for which the only
vertical orbits are great circles ($\theta = \pi/2$).
The only restriction in the present case is that 
the wheel fits inside the sphere, \ie, $r\sin\theta > a$, and that the
minimum angular velocity satisfy eq.~(\ref{eq28a}).

\subsubsection{Horizontal Orbits}

The second class of orbits is defined by $\ddot\phi = 0$, so that the angular 
velocity
is constant, say $\dot\phi = \Omega$.  From eq.~(\ref{eq23}) we see that
$\sin\beta = 0$ and hence $\beta = 0$ for these orbits, 
which implies that they are horizontal.
Then eq.~(\ref{eq24}) gives the relation between the required angular
velocity $\Omega$ and the geometrical parameters of the orbit:
\begin{equation}
\Omega^2 = {g\cot\alpha \over (2k + 1)r' + k a \cos\alpha} =
           {g\cot\alpha \over (2k + 1) r \sin\theta - (k + 1) a \cos\alpha},
\label{eq29}
\end{equation}
recalling eq.~(\ref{eq8b}).  Compare Ex.~3, sec.~244 of Routh \cite{Routh} or
sec.~407 of Milne \cite{Milne}.
  There are no steady horizontal orbits for which
$\alpha = 0$, {\it i.e.}, for which the wheel lies in the plane of the orbit.
For such an orbit the angular momentum would be constant, but the torque on
the wheel would be nonzero in contradiction.

In the following we will find that horizontal orbits are possible only for
$0 < \alpha < \pi/2.$

First, the requirement that $\Omega^2 > 0$ for real orbits puts various
restrictions on the parameters of the problem.  We examine these for the
four quadrants of angle $\alpha$.
\begin{enumerate}
\item
$0 < \alpha < \pi/2$.  Then $\cot\alpha > 0$ so we must have
\begin{equation}
r' > - {ka\cos\alpha \over 2k + 1}.
\label{eq29a}
\end{equation}
This is satisfied by all $r' > 0$ and some $r' < 0$.   However, for the
wheel to fit inside the sphere with $0 < \alpha < \pi/2$, we can have $r' < 0$
only for $\theta > \pi/2$ according to eqs.~(\ref{eq0}) and (\ref{eq8b}).
\item
$\pi/2 < \alpha < \pi$.  Then $\cos\alpha < 0$ and $\cot\alpha < 0$ so 
the numerator of (\ref{eq29}) is negative and the denominator is positive.
Hence $\Omega$ is imaginary and there are no steady orbits in this quadrant.


\item
$-\pi < \alpha < -\pi/2$.  Then $\cos\alpha < 0$ but $\cot\alpha > 0$ so 
$\Omega^2 > 0$ and $r' > 0$ and eq.~(\ref{eq29}) imposes no
to restriction.  For the wheel to fit inside the sphere with
$\alpha$ in this quadrant we must have $\theta < \pi/2$.
\item
$-\pi/2 < \alpha < 0$.  Then $\cot\alpha < 0$ so we must have
\begin{equation}
r' < - {ka\cos\alpha \over 2k + 1} < 0.
\label{eq29c}
\end{equation}
For the wheel to be inside the sphere with $r' < 0$ and $\alpha$ in this
quadrant we must have $\theta < \pi/2$.
\end{enumerate}

To obtain further restrictions on the parameters we examine under what
conditions the wheel remains in contact with the sphere.
The contact force ${\bf F}$ is deduced from eqs.~(\ref{eq2}), 
(\ref{eq17}) and  (\ref{eq21}) to be
\begin{equation}
{\bf F}/m = (-g\cos\alpha + r'\Omega^2\sin\alpha) \hat{\bf 1} -
           (g\sin\alpha + r'\Omega^2\cos\alpha) \hat{\bf 3}.
\label{eq30}
\end{equation}
It is more useful to express ${\bf F}$ in components along the $\hat{\bf r}$ and
$\hat\theta$ axes where $\hat{\bf r}$ points away from the center of the sphere
and $\hat\theta$ points towards increasing $\theta$.
The two sets of axes are related by a rotation about axis $\hat{\bf 2}$:
\begin{equation}
\hat{\bf 1} = -\cos(\theta - \alpha) \hat{\bf r} 
+ \sin(\theta - \alpha) \hat\theta,
\qquad
\hat{\bf 3} = \sin(\theta - \alpha) \hat{\bf r} 
+ \cos(\theta - \alpha) \hat\theta,
\label{eq31}
\end{equation}
so that
\begin{eqnarray}
{\bf F}/m & = & -(r'\Omega^2\sin\theta - g\cos\theta) \hat{\bf r} -
            (r'\Omega^2\cos\theta + g\sin\theta) \hat \theta
\nonumber \\
& = & -r'\Omega^2 \hat{\bf r}' + g \hat{\bf z}.
\label{eq32}
\end{eqnarray}
The second form of eq.~(\ref{eq32}) follows directly from elementary 
considerations.
The inward component of the contact force, $F_\perp = -F_r$, 
is positive and the orbits are physical provided
\begin{equation}
r'\Omega^2 > g\cot\theta.
\label{eq33}
\end{equation}

There can be no orbits with $r' < 0$ and $\theta < \pi/2$, which rules
out orbits in quadrant 4 of $\alpha$, \ie, for $-\pi/2 > \alpha < 0$.

Using eq.~(\ref{eq29}) for $\Omega^2$ in eq.~(\ref{eq33})
we deduce that contact is maintained for orbits with $r' > 0$ only if
\begin{equation}
\cot\alpha > [2k + 1 + k(a/r')\cos\alpha]\cot\theta.
\label{eq34}
\end{equation}
For $r' < 0$ the sign of the inequality is reversed.

In the third quadrant of $\alpha$ we have $\cos\alpha < 0$, so inequality
(\ref{eq34}) can be rewritten with the aid of (\ref{eq8b}) as
\begin{equation}
\cot\alpha > \left( 1 + 2k - {k \over 1 + r \sin\theta /a \abs{\cos\alpha} }
\right) \cot\theta > \cot\theta.
\label{eq34a}
\end{equation}
However, in this quadrant inequality (\ref{eq0}) tells us
\begin{equation}
\cot\alpha < \cot\left[\theta + \sin^{-1}(a/r) \right] < \cot\theta.
\label{eq34b}
\end{equation}
Hence there can be no steady orbits with $-\pi < \alpha < -\pi/2$.

Thus steady horizontal orbits are possible only for $0 < \alpha < \pi/2$.
Furthermore, since the factor in brackets of inequality (\ref{eq34}) is roughly 
2 for a wheel, this kinematic
constraint is somewhat stronger than the purely geometric relation
(\ref{eq0}).  
However, a large class of orbits remains with $\theta < \pi/2$
as well as $\theta > \pi/2$.

The coefficient of friction $\mu$ at the point of contact must be at least
$F_\parallel/F_\perp$ where $F_\parallel = \abs{F_\theta}$ from 
eq.~(\ref{eq32}).
(For $\theta > \pi/2$ and $\alpha$ near zero the tangential
friction $F_\theta$ can sometimes point in the $+\theta$ direction.)
Hence we need
\begin{equation}
\mu \geq {\abs{r'\Omega^2\cos\theta + g\sin\theta} \over r'\Omega^2\sin\theta - 
           g\cos\theta}.
\label{eq35}
\end{equation}
The acceleration of the center of mass of the wheel is $r'\Omega^2$, so
according to eq.~(\ref{eq29}) the corresponding number of $g$'s is
\begin{equation}
{\cot\alpha \over 2k + 1 + k(a/r')\cos\alpha}.
\label{eq36}
\end{equation}

Table \ref{table1} lists parameters of several horizontal orbits for
a sphere of size as might be found in a motorcycle circus.  The coefficient
of friction of rubber tires is of order one, so orbits more than a few
degrees above the equator involve very strong accelerations.  The head of
the motorcycle rider is closer to the vertical axis of the sphere than
is the center of the wheel, so the number of $g$'s experienced by the rider
is somewhat less than that given in the Table.

\begin{table}[htbp] 
\begin{center}
\parbox{5.5in}  
{\caption[ Short caption for the List of Tables. ]
{\label{table1} Parameters for horizontal circular orbits of a wheel
of radius 0.3 m rolling inside a sphere of radius 3.0 m.  The wheel
has coefficient $k = 1/2$ pertaining to its moment of
inertia.  The polar angle of the orbit is $\theta$, so orbits above the
equator of the sphere have $\theta < 90^\circ$.  The plane of the wheel
makes angle $\alpha$ to the horizontal.  The minimum coefficient of
friction required to support the motion is $\mu$.  The magnitude of the
horizontal acceleration of the center of mass is reported as the No. of
$g$'s.
}}
\vskip6pt
\begin{tabular}{ccccc}
\hline\hline
$\theta$ & $\alpha$ & $\mu$ & $v_{\rm cm}$ & No. of $g$'s \r
(deg.) & (deg.) & & (m/s) & \r
\hline
15 &  5 & 16.1 & 4.8 & 48 \r
30 &  5 & 2.82 & 8.0 & 53 \r
45 & 10 & 2.15 & 7.0 & 27 \r
60 & 10 & 1.19 & 7.9 & 27 \r
60 & 25 & 3.45 & 4.9 & 10 \r
75 & 15 & 0.96 & 6.8 & 18 \r
75 & 30 & 2.13 & 4.7 & 8  \r
90 & 25 & 0.96 & 5.3 & 10 \r
90 & 45 & 2.04 & 3.7 & 5  \r
135 & 60 & 0.56 & 2.3 & 3 \r
\hline\hline
\end{tabular}
\end{center}
\end{table}

\begin{figure}[htp]  
\postscript{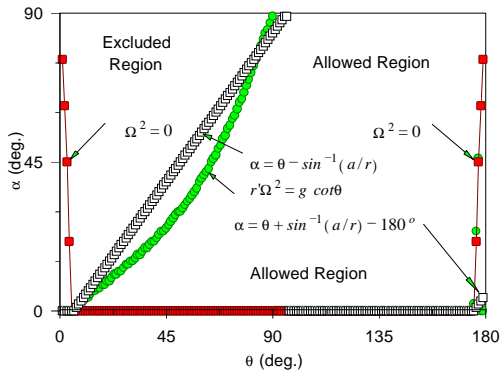}{0.4}
\begin{center}
\parbox{5.5in} 
{\caption[ Short caption for table of contents ]
{\label{alpha_wh} The allowed values of the tilt angle $\alpha$ as a function
of the angle $\theta$ of horizontal orbits for $a/r = 0.1$.  The allowed
region is bounded by three curves, derived from expressions (\ref{eq0}),
(\ref{eq29}) and (\ref{eq33}).
}}
\end{center}
\end{figure}

Figure \ref{alpha_wh} illustrates the allowed values of the tilt angle
$\alpha$ as a function of the angle $\theta$ of the plane of the orbit,
for $a/r = 0.1$ as in Table \ref{table1}.

From eq.~(\ref{eq29}) we see that $\alpha = \pi/2$, $\Omega = 0$ is a
candidate ``orbit'' in the lower hemisphere.  On such an ``orbit'' the wheel
is standing vertically at rest, and is not stable against falling over.
We infer that stability will only occur for $\Omega$ greater than
some minimum value not revealed by the analysis thus far.  

\subsection{Stability Analysis}

A completely general analysis of the stability of the steady circular orbits
found above appears to be very difficult.  We give a fairly general analysis
for vertical orbits, but for horizontal orbits we obtain results only
for orbits with $\theta = \pi/2$, {\it i.e.}, orbits about the equator of
the sphere, and for orbits of ``small'' wheels.

We follow the approach of sec.~405
of Milne \cite{Milne} where it was shown how the steady motion of a disk
rolling in a straight line on a horizontal plane is stable if the angular
velocity is great enough.  It was also shown that the small oscillatory
departures from steady motion lead to an oscillatory path of the point of
contact of the wheel with the plane.  Hence in the present case we must 
consider perturbations
that carry the wheel away from the plane of the steady orbit.

The difficulty is that there are in general four degrees of freedom for
departures from steady motion: the axis of the wheel can be perturbed in
two directions and the angular velocity $\dot\phi$ can be perturbed as well as
the angle $\theta$ to the point of contact.  However, the procedure to
eliminate the unknown force of contact from the six equations of motion
of a rigid body leaves only three equations of motion.
We will obtain solutions to the perturbed equations of motions only in
special cases where there are in effect just two or three degrees of freedom.
A more general analysis might be possible using the contact force found in 
steady motion as a first approximation to the contact force in perturbed motion,
but we do not pursue this here.

A wheel
rolling with a steady circular orbit on a plane can suffer only three types of
perturbations and the results of an
analysis are reported in Ex.~3, sec.~244 of Routh \cite{Routh}.  For a
sphere rolling within a fixed sphere the direction of what we call axis 3 always
points to the center of the fixed sphere so there are only two perturbations to 
consider and the solution is relatively straightforward, as reviewed in
sec.~4 below.  The stability of 
horizontal orbits of rolling spheres lends confidence that stable orbits
also exist for wheels.

\subsubsection{Vertical Orbits}

\begin{figure}[htp]  
\postscript{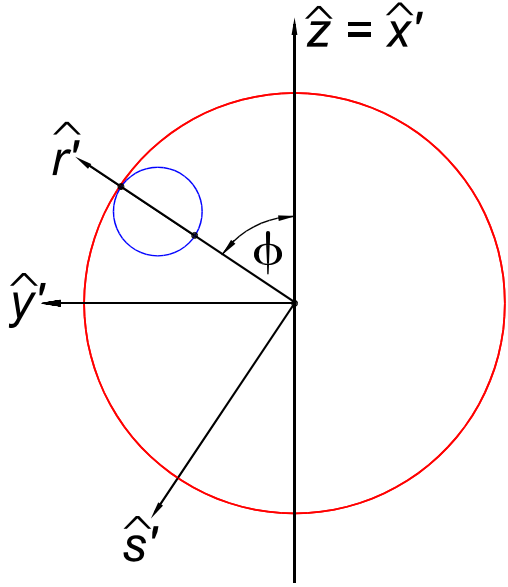}{0.4}
\begin{center}
\parbox{5.5in} 
{\caption[ Short caption for table of contents ]
{\label{fig5} For vertical orbits the $x'$-axis is identical with the $z$ axis.
The axis $\hat{\bf s}' = \hat{\bf z}' \times \hat{\bf r}'$ is in the direction of
the unperturbed $\hat{\bf 2}$-axis.
}}
\end{center}
\end{figure}

We define the $(x',y',z')$ coordinate system to have the $x'$-axis vertical:
$\hat{\bf x}' = \hat{\bf z}$.  In steady motion we have
\begin{equation}
\alpha = 0, \qquad \hat{\bf 1} = -\hat{\bf z}', \qquad \mbox{and} \qquad
\hat{\bf 3} = \hat{\bf r}' = \hat{\bf x}' \cos\phi + \hat{\bf y}' \sin\phi,
\label{eq60}
\end{equation}
where $\phi$ is the azimuth of the center of the wheel from the 
$\hat{\bf x}'$-axis.  Thus $\phi = 0$ at the top of the orbit.
To discuss departures from steady motion in which the $\hat{\bf 1}$-axis is no
longer parallel to the $\hat{\bf z}'$-axis, it is useful to have a unit triad 
$(\hat{\bf r}',\hat{\bf s}',\hat{\bf z}')$ defined by eq.~(\ref{eq60}) and
\begin{equation}
\hat{\bf s}' = \hat{\bf z}' \times \hat{\bf r}' 
= -\hat{\bf x}' \sin\phi + \hat{\bf y}' \cos\phi,
\label{eq61}
\end{equation}
with $\phi$ defined as before. See Fig.~\ref{fig5}.
  The surface of the sphere at the point of
contact is parallel to the $s'$-$z'$ plane.  Axes $\hat{\bf r}'$ and $\hat{\bf z}'$
rotate about the $z'$-axis with angular velocity $\dot\phi$, so that
\begin{equation}
{d\hat{\bf r}' \over dt} = \dot\phi \hat{\bf s}', \qquad \mbox{and} \qquad
{d\hat{\bf s}' \over dt} = -\dot\phi \hat{\bf r}'.
\label{eq61a}
\end{equation}

The perturbed $\hat{\bf 1}$-axis can then be written
\begin{equation}
\hat{\bf 1} = \epsilon_r \hat{\bf r}' + \epsilon_s \hat{\bf s}' - \hat{\bf z}',
\qquad \mbox{with}
\qquad \abs{\epsilon_r},\abs{\epsilon_s} \ll 1,
\label{eq62}
\end{equation}
where throughout the stability analysis we ignore second-order terms.
Writing
\begin{equation}
\hat{\bf 3} = \hat{\bf r}' + \delta_s \hat{\bf s}' + \delta_z \hat{\bf z}', 
\qquad \mbox{with}
\qquad \abs{\delta_s},\abs{\delta_z} \ll 1,
\label{eq63}
\end{equation}
the condition $\hat{\bf 1} \cdot \hat{\bf 3} = 0$ requires that $\delta_z =
\epsilon_r$.
Then to first order,
\begin{equation}
\hat{\bf 2} = \hat{\bf 3} \times \hat{\bf 1} 
= - \delta_s \hat{\bf r}' + \hat{\bf s}' + \epsilon_s \hat{\bf z}'.
\label{eq64}
\end{equation}
We expect that vector $\hat{\bf 2}$ will remain parallel to the surface of the
sphere even for large departure from steady motion, so $\hat{\bf 2}$ must remain 
in the $s'$-$z'$ plane.  Hence, $\delta_s = 0$, and
\begin{equation}
\hat{\bf 3} = \hat{\bf r}' + \epsilon_r \hat{\bf z}'.
\label{eq65}
\end{equation}
Also, we can identify $\alpha$ as the tilt angle of the $\hat{\bf 3}$-axis to the
$r'$-$s'$ plane, so that
\begin{equation}
\alpha = \epsilon_r.
\label{eq66}
\end{equation}

The analysis proceeds along the lines of sec.~2.1 except that now we express
all vectors in terms of the triad $(\hat{\bf r}',\hat{\bf s}',\hat{\bf z}')$.  
To the
first approximation the angular velocity of the wheel about the $\hat{\bf 1}$-axis
is
still given by $\omega_1 = (r'/a)\dot\phi$.  From eqs.~(\ref{eq61a}) and 
(\ref{eq62}) we find
\begin{equation}
{d \hat{\bf 1} \over dt} = (\dot\epsilon_r - \epsilon_s \dot\phi) \hat{\bf r}' +
                      (\epsilon_r \dot\phi + \dot\epsilon_s) \hat{\bf s}',
\label{eq67}
\end{equation}
\begin{equation}
\hat{\bf 1} \times {d \hat{\bf 1} \over dt} = 
       (\epsilon_r \dot\phi + \dot\epsilon_s) \hat{\bf r}' -                      
       (\dot\epsilon_r - \epsilon_s \dot\phi) \hat{\bf s}',
\label{eq68}
\end{equation}
so that eq.~(\ref{eq7}) yields
\begin{equation}
\vec\omega = \omega_1 \hat{\bf 1} + \hat{\bf 1} \times {d\hat{\bf 1} \over dt} = 
[(1 + r'/a) \epsilon_r \dot\phi + \dot\epsilon_s] \hat{\bf r}'
           - [\dot\epsilon_r - (1 + r'/a) \epsilon_s \dot\phi] \hat{\bf s}'
           - (r'/a)\dot\phi \hat{\bf z}'.
\label{eq69}
\end{equation}
Then eq.~(\ref{eq7b}) tells us
\begin{equation}
{{\bf L} \over ma} = 2ka\omega_1 \hat{\bf 1} + ka \hat{\bf 1} \times {d\hat{\bf 1}
 \over dt}
          = k[(2r' + a) \epsilon_r \dot\phi + a\dot\epsilon_s] \hat{\bf r}'
           - k[a\dot\epsilon_r - (2r' + a) \epsilon_s \dot\phi] \hat{\bf s}'
           - 2kr'\dot\phi \hat{\bf z}',
\label{eq70}
\end{equation}
so that to first order of smallness
\begin{eqnarray}
{1 \over ma}{d{\bf L} \over dt} & = & 
  k [ 2(r' + a) \dot\epsilon_r \dot\phi 
     + (2r' + a) (\epsilon_r \ddot\phi - \epsilon_s \dot\phi^2)
     + a \ddot\epsilon_s ] \hat{\bf r}'   \nonumber \\
  & & - k[ a \ddot\epsilon_r  
          - (2r' + a) (\epsilon_r \dot\phi^2 - \epsilon_s \ddot\phi)
          - 2(r' + a) \dot\epsilon_s \dot\phi ] \hat{\bf s}' \nonumber \\
  & & - 2k(r' \ddot\phi + \dot r' \dot\phi) \hat{\bf z}'.
\label{eq71}  
\end{eqnarray}
In this we have noted from eq.~(\ref{eq8b}) that $\dot r' = 
r\dot\theta \sin\theta$ to first order, and that $\dot\theta$ is small. 
Next,
\begin{equation}
{d{\bf r}_{\rm cm} \over dt} = a \hat{\bf 3} \times \vec\omega 
       \approx a \hat{\bf r}' \times \vec\omega = r' \dot\phi \hat{\bf s}' 
     - [a \dot\epsilon_r - (r' + a) \epsilon_s \dot\phi] \hat{\bf z}'.
\label{eq72}
\end{equation}
Then to first order,
\begin{equation}
{d^2 {\bf r}_{\rm cm} \over dt^2} = - r' \dot\phi^2 \hat{\bf r}'
     + (r' \ddot\phi + \dot r' \dot\phi) \hat{\bf s}' 
     - [a \ddot \epsilon_r 
     - (r' + a) (\dot\epsilon_s \dot\phi + \epsilon_s \ddot\phi) ] \hat{\bf z}',
\label{eq73}
\end{equation}
so that
\begin{eqnarray}
\hat{\bf 3} \times {d^2 {\bf r}_{\rm cm} \over dt^2} & = &
(\hat r' + \epsilon_r \hat{\bf z}') \times {d^2 {\bf r}_{\rm cm} \over dt^2} = 
\nonumber \\
  & & - r' \epsilon_r \ddot\phi \hat{\bf r}'
      + [ a \ddot \epsilon_r - r' \epsilon_r \dot\phi^2 
         - (r' + a) (\dot\epsilon_s \dot\phi + \epsilon_s \ddot\phi)] \hat{\bf s}'
      + (r' \ddot\phi + \dot r' \dot\phi) \hat{\bf z}' 
\label{eq74}
\end{eqnarray}
Also,
\begin{equation}
\hat{\bf 3} \times \hat{\bf z} = (\hat{\bf r}' + \epsilon_r \hat{\bf z}') \times 
       (\cos\phi \hat{\bf r}' - \sin\phi \hat{\bf s}') =
       \epsilon_r \sin\phi \hat{\bf r}' + \epsilon_s \cos\phi \hat{\bf s}'
     - \sin\phi \hat{\bf z}'.
\label{eq75}
\end{equation}

The $r'$, $s'$ and $z'$ components of the equation of motion (\ref{eq4}) are 
then
\begin{equation}
0 = [(2k + 1)r' + ka] \epsilon_r \ddot\phi + 2k(r' + a) \dot\epsilon_r \dot\phi 
     - g \epsilon_r \sin\phi
     - k(2r' + a) \epsilon_s \dot\phi^2 + ka \ddot\epsilon_s,
\label{eq76}
\end{equation}
\begin{eqnarray}
0 & = &  [(2k + 1)r + ka] \epsilon_r \dot\phi^2 
    - (k + 1)a \ddot\epsilon_r - g \epsilon_r \cos\phi \nonumber \\
 & & + (2k + 1)(r' + a) \dot\epsilon_s \dot\phi
    + [(2k + 1)r' + (k + 1)a] \epsilon_s \ddot\phi,
\label{eq77}
\end{eqnarray}
\begin{equation}
0 = (2k + 1)(r' \ddot\phi + \dot r' \dot\phi) - g \sin\phi.
\label{eq78}
\end{equation}
If the perturbations $\epsilon_r$, $\epsilon_s$ and $\dot r'$ are set to
zero eqs.~(\ref{eq76}) and (\ref{eq77}) become trivial while eq.~(\ref{eq78})
becomes the steady equation of motion (\ref{eq26}).

The general difficulty with this analysis is that there are only three
equations, (\ref{eq76}-\ref{eq78}), while there are four perturbations,
$\epsilon_r$, $\epsilon_s$, $\ddot\phi$ and $\dot\theta$.
The perturbation $\dot\theta$ appears only
in eq.~(\ref{eq78}) via $\dot r'$; its effect on $r'$ leads only to second-order
terms in eqs.~(\ref{eq76}-\ref{eq77}).  If we could neglect the terms in 
$\ddot\phi$ in eqs.~(\ref{eq76}-\ref{eq77}) then these two equations would
describe only the perturbations $\epsilon_r$ and $\epsilon_s$ to first order
and a solution could be completed.

Therefore we restrict our attention to the top of the orbit, $\phi = 0$,
where eq.~(\ref{eq78}) tells us that $\ddot\phi = 0$ to leading order.
The angular velocity $\dot\phi_0$ at this point is a minimum so the
gyroscopic stability of the wheel is the least here.  Hence if the orbit
is stable at $\phi = 0$ it will be stable at all $\phi$.

The forms of eqs.~(\ref{eq76}) and (\ref{eq77}) for $\phi = 0$ indicate that 
if $\epsilon_r$
and $\epsilon_s$ are oscillatory then they are $90^\circ$ out of phase.
Therefore we seek solutions
\begin{equation}
\epsilon_r = \epsilon_r \cos\omega t, \qquad
\epsilon_s = \epsilon_s \sin\omega t, 
\label{eq100}
\end{equation}
where $\omega$ now represents the oscillation frequency.  The coupled
equations of motion then yield the simultaneous linear equations
\begin{eqnarray}
2k(r' + a) \dot\phi_0 \omega \epsilon_r 
  & + & [ka \omega^2 + k(2r' + a) \dot\phi_0^2] \epsilon_s = 0 \nonumber \\
\{(k + 1)a \omega^2 + [(2k + 1)r' + ka] \dot\phi_0^2 - g\} \epsilon_r
  & + & (2k + 1)(r' + a) \dot\phi_0 \omega \epsilon_s = 0.
\label{eq102}
\end{eqnarray}
These equations are consistent only if the determinant of the coefficient
matrix vanishes, which leads to the quadratic equation
\begin{equation}
A \omega^4 - B \omega^2 - C = 0,
\label{eq103}
\end{equation}
with solutions
\begin{equation}
\omega^2 = {B \pm \sqrt{B^2 + 4AC} \over 2A},
\label{eq103a}
\end{equation}
where
\begin{equation}
A = k(k + 1)a^2,
\label{eq104}
\end{equation}
\begin{equation}
B = kag + k[(2k + 1)(2r'^2 + a^2) + (4k + 1)ar'] \dot\phi_0^2,
\end{equation}
and
\begin{equation}
C = k(2r' + a) \left([(2k+1)r' + ka] \dot\phi_0^2  - g \right) \dot\phi_0^2.
\label{eq105}
\end{equation}

Since $A$ and $B$ are positive there are real, positive roots whenever 
$B^2 + 4AC$ is positive, {\it i.e.}, for $C > -B^2/4A$.  In particular, this is
satisfied for positive $C$, or equivalently for
\begin{equation}
\dot\phi_0^2 > {g \over (2k + 1)r' + ka}.
\label{eq106}
\end{equation}
However, this is less restrictive than the elementary result (\ref{eq28a}) 
that the wheel stay in contact with the sphere!  All vertical orbits for
which the wheel remains in contact with the sphere are stable against small
perturbations.

The stability analysis yields the formal result that if $(\phi,\dot\phi) = 
(0,0)$ then $\omega = \sqrt{g/(k+1)a}$.  We recognize this as the frequency of
oscillation  of a
simple pendulum formed by suspending the wheel from a point on its rim,
the motion being perpendicular to the plane of the wheel.

\subsubsection{Horizontal Orbits}

We expect the stability analysis of horizontal orbits to be nontrivial
since we have identified steady orbits that are ``obviously'' unstable.

The spirit of the analysis has been set forth in the preceding sections.
For horizontal orbits the $(x',y',z')$ coordinate system can be taken as
identical with the $(x,y,z)$ system, so we drop symbol $'$ in this section.
We introduce a triad $(\hat{\bf r},\hat{\bf s},\hat{\bf z})$ with $\hat r$ being
the perpendicular unit vector from the $z$-axis toward the center of the wheel.
Then $\hat{\bf s}$ points in the direction of the motion of the center of the
wheel in case of steady motion.

\begin{figure}[htp]  
\postscript{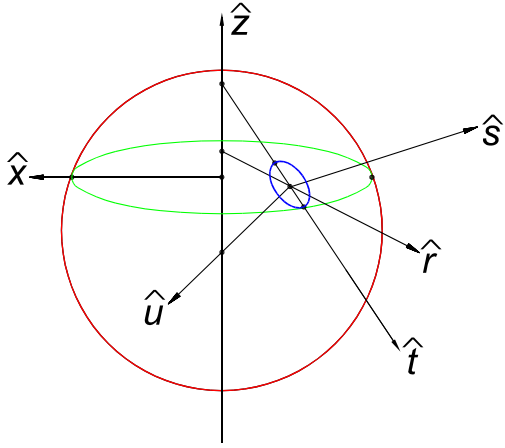}{0.4}
\begin{center}
\parbox{5.5in} 
{\caption[ Short caption for table of contents ]
{\label{fig6} For horizontal orbits of a wheel rolling inside a sphere
the $(x,y,z)$ axes are identical with the $(x',y',z')$ axes. 
The $\hat{\bf r}$-$\hat{\bf s}$ plane is horizontal. 
The axes $\hat{\bf u}$, $\hat{\bf s}$ and $\hat{\bf t}$ are along the unperturbed
directions
of the $\hat{\bf 1}$, $\hat{\bf 2}$ and $\hat{\bf 3}$ axes, respectively.
Axes $\hat{\bf t}$ and $\hat{\bf u}$ lie in the vertical plane 
$\hat{\bf r}$-$\hat{\bf z}$.
}}
\end{center}
\end{figure}

It is also useful to introduce a unit triad that points along the 
$(\hat{\bf 1},\hat{\bf 2},\hat{\bf 3})$ axes for steady motion.  The $\hat{\bf s}$
axis already
points along the $\hat{\bf 2}$ axis for steady motion, so we only need define
$\hat{\bf t}$ as being along the direction of $\hat{\bf 3}$, and $\hat{\bf u}$ as
being
along the direction of $\hat{\bf 1}$ for steady motion, as shown in 
Fig.~\ref{fig6}.
Then, 
$(\hat{\bf s},\hat{\bf t},\hat{\bf u})$ form a right-handed unit triad.  The vertical, 
$\hat{\bf z}$, is then related by
\begin{equation}
\hat{\bf z} = -\sin\alpha_0 \hat{\bf t} - \cos\alpha_0 \hat{\bf u},
\label{eq111}
\end{equation}
where $\alpha_0$ is the angle of inclination of the wheel to the horizontal
in steady motion.  The triad $(\hat{\bf s},\hat{\bf t},\hat{\bf u})$ rotates about
the $\hat{\bf z}$-axis with angular velocity $\dot\phi$, so that
\begin{equation}
{d\hat{\bf s} \over dt} = \dot\phi \hat{\bf z} \times \hat{\bf s} 
= - \dot\phi \cos\alpha_0 \hat{\bf t} + \dot\phi \sin\alpha_0 \hat{\bf u},
\label{eq112}
\end{equation}
\begin{equation}
{d\hat{\bf t} \over dt} = \dot\phi \cos\alpha_0 \hat{\bf s},
\label{eq113}
\end{equation}
and
\begin{equation}
{d\hat{\bf u} \over dt} = - \dot\phi \sin\alpha_0 \hat{\bf s}.
\label{eq114}
\end{equation}

We now consider small departures from steady motion.  The $\hat{\bf 1}$-axis
deviates slightly from the $\hat{\bf u}$-axis according to
\begin{equation}
\hat{\bf 1} = \epsilon_s \hat{\bf s} + \epsilon_t \hat{\bf t} + \hat{\bf u}, 
\qquad
\abs{\epsilon_s},\abs{\epsilon_t} \ll 1.
\label{eq115}
\end{equation}
The $\hat{\bf 3}$-axis departs slightly from the $t$-axis, but to the first 
approximation it remains in a vertical plane, \ie, the $t$-$u$ plane.
Then we have
\begin{equation}
\hat{\bf 2} = \hat{\bf s} - \epsilon_s \hat{\bf u}, \qquad \mbox{and} \qquad
\hat{\bf 3} = \hat{\bf t} - \epsilon_t \hat{\bf u}.
\label{eq116}
\end{equation}
With the above definitions the signs of angles $\alpha$ and $\epsilon_t$
are opposite:
\begin{equation}
\Delta\alpha = - \epsilon_t, \qquad \dot\alpha = -\dot\epsilon_t.
\label{eq117}
\end{equation}

To first approximation the component $\omega_1$ of the angular velocity of
the wheel about its axis remains $\omega_1 = (r'/a) \dot\phi$.  Then
\begin{equation}
{d\hat{\bf 1} \over dt} = 
                (- \dot\phi \sin\alpha_0 + \dot\epsilon_s  
                 - \epsilon_t \dot\phi \cos\alpha_0) \hat{\bf s}
              - (\epsilon_s \dot\phi \cos\alpha_0 - \dot\epsilon_t) \hat{\bf t} 
                     + \epsilon_s \dot\phi \sin\alpha_0 \hat{\bf u},
\label{eq118}
\end{equation}
\begin{equation}
\hat{\bf 1} \times {d\hat{\bf 1} \over dt} = 
     (\epsilon_s \dot\phi \cos\alpha_0 - \dot\epsilon_t) \hat{\bf s}
   - (\dot\phi \sin\alpha_0 + \dot\epsilon_s 
      - \epsilon_t \dot\phi \cos\alpha_0) \hat{\bf t}
   + \epsilon_t \dot\phi \sin\alpha_0 \hat{\bf u},
\label{eq119}
\end{equation}
so that
\begin{eqnarray}
\vec\omega & = & \omega_1 \hat{\bf 1} + \hat{\bf 1} \times {d\hat{\bf 1} \over dt} \nonumber \\
    & = & [(r'/a + \cos\alpha_0)\epsilon_s \dot\phi - \dot\epsilon_t] \hat{\bf s}
\nonumber \\
    & &  - [\dot\phi \sin\alpha_0 - \dot\epsilon_s 
         - (r'/a + \cos\alpha_0) \epsilon_t \dot\phi] \hat{\bf t} \nonumber \\
    & &  + (r'/a + \epsilon_t \sin\alpha_0) \dot\phi \hat{\bf u}, 
\label{eq120} 
\end{eqnarray}
and
\begin{eqnarray}
{{\bf L} \over ma} 
 & = & 2ka \omega_1 \hat{\bf 1} + ka \hat{\bf 1} \times {d\hat{\bf 1} \over dt} \nonumber \\
 & = & k [(2r' + a \cos\alpha_0) \epsilon_s \dot\phi - a \dot\epsilon_t] 
\hat{\bf s}
\nonumber \\
 & & - k [a \dot\phi \sin\alpha_0 - a \dot\epsilon_s
        - (2r' + a \cos\alpha_0) \epsilon_t \dot\phi] \hat{\bf t} \nonumber \\
 & &  + k (2r' + a \epsilon_t \sin\alpha_0) \dot\phi \hat{\bf u}.
\label{eq121} 
\end{eqnarray}     
Then to the first approximation
\begin{eqnarray}
{1 \over ma}{d {\bf L} \over dt} 
  & = & -k[(2r' + a \cos\alpha_0) \dot\phi^2 \sin\alpha_0 
          - 2(r' + a \cos\alpha_0) \dot\epsilon_s \dot\phi) \nonumber \\
  & & \phantom{aaaaaaa}
          - (2r' \cos\alpha_0 + a \cos 2\alpha_0) \epsilon_t \dot\phi^2
          + a \ddot\epsilon_t] \hat{\bf s}     \nonumber \\
  & & -k[a \ddot\phi \sin\alpha_0 
         + (2r' + a \cos\alpha_0) \epsilon_s \dot\phi^2 \cos\alpha_0 
         - a \ddot\epsilon_s 
         - 2(r' + a \cos\alpha_0) \dot\epsilon_t \dot\phi)] \hat{\bf t} 
\nonumber \\
  & & + k[2r' \ddot\phi + 2\dot r' \dot\phi 
        + (2r' + a \cos\alpha_0) \epsilon_s \dot\phi^2 \sin\alpha_0] \hat{\bf  u}.
\label{eq122} 
\end{eqnarray}
Unlike the case of vertical orbits, for horizontal orbits the factor 
$\ddot\phi$ has no zeroeth-order
component and we neglect terms like $\epsilon \ddot\phi$.

Similarly
\begin{eqnarray}
{d {\bf r}_{\rm cm} \over dt} & = & a \hat{\bf 3} \times \vec\omega =
                 a (\hat{\bf t} - \epsilon_t \hat{\bf u}) \times \vec\omega 
\nonumber \\
  & = & r' \dot\phi \hat s
     - [(r' + a \cos\alpha_0) \epsilon_s \dot\phi - a \dot\epsilon_t] \hat{\bf u}, 
\label{eq123} 
\end{eqnarray}
\begin{eqnarray}
{d^2 {\bf r}_{\rm cm} \over dt^2} & = & 
 [r' \ddot\phi + \dot r' \dot\phi 
  + (r' + a \cos\alpha_0) \epsilon_s \dot\phi^2 \sin\alpha_0 
  - a \dot\epsilon_t \dot\phi \sin\alpha_0] \hat{\bf s}     \nonumber \\
 & & - r' \dot\phi^2 \cos\alpha_0 \hat{\bf t} 
+  [r' \dot\phi^2 \sin\alpha_0 - (r' + a \cos\alpha_0) \dot\epsilon_s \dot\phi 
    + a \ddot\epsilon_t ] \hat{\bf u},
\label{eq124}
\end{eqnarray}
and
\begin{eqnarray}
\hat{\bf 3} \times {d^2 {\bf r}_{\rm cm} \over dt^2} & = & 
        (\hat{\bf t} - \epsilon_t \hat{\bf u}) \times {d^2 {\bf r}_{\rm cm} \over
 dt^2} 
        \nonumber \\
 & = & [r' \dot\phi^2 \sin\alpha_0 
        - (r' + a \cos\alpha_0) \dot\epsilon_s \dot\phi 
        - r' \epsilon_t \dot\phi^2 \cos\alpha_0 + a \ddot\epsilon_t] \hat{\bf s}
   \nonumber  \\
 & & - [r' \ddot\phi + \dot r' \dot\phi 
  + (r' + a \cos\alpha) \epsilon_s \dot\phi^2 \sin\alpha_0 
  - a \dot\epsilon_t \dot\phi \sin\alpha_0] \hat{\bf u}.
\label{eq125} 
\end{eqnarray}

We also need
\begin{equation}
\hat{\bf 3} \times \hat{\bf z}  =  
(\hat{\bf t} - \epsilon_t \hat{\bf u}) \times (-\sin\alpha_0 \hat{\bf t} 
- \cos\alpha_0 \hat{\bf u})
= - (\cos\alpha_0 + \epsilon_t \sin\alpha_0) \hat{\bf s}.
\label{eq126}
\end{equation}

The $s$, and $t$ and $u$ components of the equation of motion (\ref{eq4}) are 
\begin{eqnarray}
0 & = & [(2k+1)r' + ka \cos\alpha_0] \dot\phi^2 \sin\alpha_0 - g \cos\alpha_0
          - (2k+1)(r' + a \cos\alpha_0] \dot\epsilon_s \dot\phi
          \nonumber \\
  & & - [(2k+1)r' \cos\alpha_0 + ka \cos 2\alpha_0] \epsilon_t \dot\phi^2 
      - g \epsilon_t \sin\alpha_0 + (k+1) a \ddot\epsilon_t,
\label{eq127}
\end{eqnarray}    
\begin{equation}
0 =  ka \ddot\phi \sin\alpha_0 
    + k (2r' + a \cos\alpha_0) \epsilon_s \dot\phi^2 \sin\alpha_0 
    - ka \ddot\epsilon_s  - 2k (r' + a \cos\alpha_0)\dot\epsilon_t \dot\phi,
\label{eq128}
\end{equation}
and
\begin{equation}
0 = (2k+1)(r' \ddot\phi + \dot r' \dot\phi) 
    + [(2k+1)r' + (k+1)a \cos\alpha_0] \epsilon_s \dot\phi^2 \sin\alpha_0
    - a \dot\epsilon_t \dot\phi \sin\alpha_0.
\label{eq129}
\end{equation}
The leading terms of these three equations are just 
eqs.~(\ref{eq23})-(\ref{eq25}) for $\beta = 0$.  Therefore we can write
$\dot\phi = \Omega + \dot\delta$ where $\Omega$ is the angular velocity
of the steady horizontal orbit and $\delta$ is a small correction.

Although the derivative of $r'$,
\begin{equation}
\dot r' = r \dot\theta \cos\theta_0 + a \dot\alpha \sin\alpha_0 =
          r \dot\theta \cos \theta_0 - a \dot\epsilon_t \sin\alpha_0,
\label{eq130}
\end{equation}
appears only in eq.~(\ref{eq129}), in general the perturbation $\dot\theta$ is 
not decoupled from $\epsilon_s$ and $\epsilon_t$ as was the case for
vertical orbits.  Thus far, we have found a way to proceed only in somewhat
special cases in which the $\theta$ perturbation can be ignored, as
described in secs.~2.3.3 and 2.3.4.

\subsubsection{Orbits Near the Equator}

It appears possible to carry the analysis forward 
for the special case $\theta_0 = \pi/2$, the orbit on the equator of the sphere.
This case is, however, of interest.

Assuming $\theta_0 = \pi/2$ the equations of motion (\ref{eq127}-\ref{eq129})
then provide three relations among the three perturbations $\delta$,
$\epsilon_s$ and $\epsilon_t$.  For this we consider only the first-order
terms, noting that $\dot\phi^2 \approx \Omega^2 + 2\Omega\dot\delta$ and
\begin{equation}
r' = r\sin\theta - a\cos\alpha \approx r'_0 + r \Delta\theta \cos\theta_0 +
a \Delta\alpha \sin\alpha_0 = r'_0 - a \epsilon_t \sin\alpha_0,
\label{eq131}
\end{equation}
where $r_0 = r - a\cos\alpha_0$
for $\theta_0 = \pi/2$, recalling eq.~(\ref{eq117}).
Also, from the form of eqs.~(\ref{eq127}-\ref{eq129}) we infer that if
the perturbations are oscillatory then $\delta$ and $\epsilon_s$ have the
same phase which is $90^\circ$ from that of $\epsilon_t$.  Therefore we
seek solutions of the form
\begin{equation}
\delta = \delta\sin\omega t, \qquad
\epsilon_s = \epsilon_s \sin\omega t, \qquad \mbox{and} \qquad
\epsilon_t = \epsilon_t \cos\omega t,
\label{eq132}
\end{equation}
where $\omega$ is the frequency of oscillation.  The first-order terms of the
differential equations (\ref{eq127}-\ref{eq129}) then yield the
algebraic relations
\begin{eqnarray}
0 & = & 2 \Omega \sin\alpha_0 [(2k+1)r'_0 + ka \cos\alpha_0] \omega \delta
        - \Omega (2k+1) (r'_0 + a \cos\alpha_0) \omega \epsilon_s
\nonumber \\
  & & - \{ \Omega^2 [ (2k+1)r'_0 \cos\alpha_0 + (k + \sin^2\alpha_0) a ] 
           - g \sin\alpha_0 + (k + 1) a \omega^2 \} \epsilon_t,
\nonumber \\
0 & = & -ka \sin\alpha_0 \omega^2 \delta 
    + [k \Omega^2 \cos\alpha_0 (2r'_0 + a \cos\alpha_0) + ka \omega^2 ] 
       \epsilon_s
    + 2k \Omega (r'_0 + a \cos\alpha_0) \omega \epsilon_t, 
\label{eq133} \\
0 & = & -(2k+1) r'_0 \omega^2 \delta
     + \Omega^2 \sin\alpha_0 [(2k+1)r'_0 + (k+1)a \cos\alpha_0] \epsilon_s
     + 2(k + 1) \Omega \sin\alpha_0 a \omega \epsilon_t. 
\nonumber
\end{eqnarray}
These equations have the form
\begin{equation}
\begin{array}{ccccccc}
A_{11} \omega \delta & + & A_{12} \omega \epsilon_s & + & 
   (A_{13} + B_{13} \omega^2) \epsilon_t    & = & 0 \\
A_{21} \omega^2 \delta & + & (A_{22} + B_{22} \omega^2) \epsilon_s & + & 
    A_{23} \omega \epsilon_t & = & 0 \\
A_{31} \omega^2 \delta & + & A_{32} \epsilon_s & + & A_{33} \omega \epsilon_t
   & = & 0 \\
\end{array}
\label{eq134}
\end{equation}
To have consistency the determinant of the coefficient matrix must 
vanish, which leads quickly to the quadratic equation
\begin{equation}
A \omega^4 - B \omega^2 - C = 0,
\label{eq135}
\end{equation}
where
\begin{equation}
A = B_{13}B_{22}A_{31},
\label{eq136}
\end{equation}
\begin{equation}
B = A_{11}B_{22}A_{33} + A_{12}A_{23}A_{31} + B_{13}A_{21}A_{32} 
  - A_{13}B_{22}A_{31} - B_{13}A_{22}A_{31} - A_{12}A_{21}A_{33},
\label{eq137}
\end{equation}
and
\begin{equation}
C = A_{11}A_{22}A_{33} + A_{13}A_{21}A_{32}
  - A_{13}A_{22}A_{31} - A_{11}A_{23}A_{32}. 
\label{eq138}
\end{equation}

From numerical evaluation it appears that $A$, $B$ and $C$ are all positive
for angular velocities $\Omega$ that obey eq.~(\ref{eq29}).  That is, all
steady orbits at the equator of the sphere are stable.  There is both a fast
and slow oscillation about steady motion for these orbits, an effect
familiar from nutations of a symmetric top.

\subsubsection{Small Wheel Inside a Large Sphere}

The analysis can also 
be carried further in the approximation that the radius $a$ of
the wheel is much less than the radius $r$ of the fixed sphere.  In this
case the perturbation in angle $\theta$ of the orbit is of higher
order than the perturbations in azimuth $\phi$ and in the angles $\epsilon_s$
and $\epsilon_t$ related to the axes of the wheel.  A solution describing
the three first-order perturbations can then be obtained.

For the greatest simplification we also require that
\begin{equation}
a \ll r'_0 \approx r\sin\theta_0.
\label{eq201}
\end{equation}
Thus we restrict our attention to orbits significantly different from the
special cases of motion near the poles of the fixed sphere.

In the present approximation the first-order terms of the perturbed equations
of motion (\ref{eq127}-\ref{eq129}) are
\begin{equation}
2 \Omega \dot\delta \sin\alpha_0 = \Omega \dot\epsilon_s 
+ \left( \Omega^2 \cos\alpha_0 +  {g \sin\alpha_0 \over (2k + 1) r'_0} \right)
\epsilon_t,
\label{eq202}
\end{equation}    
\begin{equation}
\epsilon_s = {\dot\epsilon_t \over \Omega \sin\alpha_0},
\label{eq203}
\end{equation}
and
\begin{equation}
\ddot\delta = - \epsilon_s \Omega^2 \sin\alpha_0.
\label{eq204}
\end{equation}
Inserting (\ref{eq203}) into (\ref{eq204}) we can integrate the latter to find
\begin{equation}
\dot\delta = - \Omega \epsilon_t.
\label{eq205}
\end{equation}
Using this and the derivative of (\ref{eq203}) in (\ref{eq202}) we find that
$\epsilon_t$ obeys
\begin{equation}
\ddot\epsilon_t + \left[ \Omega^2 \sin\alpha_0 (\cos\alpha_0 + 2 \sin\alpha_0)
 +  {g \sin^2\alpha_0 \over (2k + 1) r'_0} \right] \epsilon_t = 0.
\label{eq206}
\end{equation}    
The the frequency $\omega$ of the perturbations is given by 
\begin{equation}
\omega^2 = \Omega^2 \sin\alpha_0 (\cos\alpha_0 +2 \sin\alpha_0)
 +  {g \sin^2\alpha_0 \over (2k + 1) r'_0} = \Omega^2 \tan\alpha_0
(1 + \sin 2\alpha_0),
\label{eq207}
\end{equation}    
using eqs.~(\ref{eq29}) and (\ref{eq201}). 

Thus all orbits for small wheels are stable if condition (\ref{eq201}) holds.
We conjecture that orbits for large wheels are also stable if (\ref{eq201}) is
satisfied.

For steady orbits that lie very near the poles,
\ie, those that have $r'_0 \lsim a$, we conjecture that the motion is stable
only for $\Omega$ greater than some minimum value. 
For a wheel spinning about its axis on a horizontal plane the stability
condition is
\begin{equation}
\Omega^2 > {g \over (2k+1)a}.
\label{eq208}
\end{equation}
See, for example, sec.~55 of Deimel \cite{Deimel}.
However, we have been unable to deduce the generalization of this  constraint
to include the dependence on $r$ and $\theta_0$ for small $r \sin\theta_0$.

\section{Wheel Rolling Outside a Fixed Sphere}

Equations (\ref{eq00})-(\ref{eq25}) hold for a wheel rolling outside a
sphere as well as inside when the geometric relation (\ref{eq0})
is rewritten as
\begin{equation}
\theta < \alpha < \pi + \theta.
\label{eq81}
\end{equation}

We expect no vertical orbits as the wheel will lose contact with the sphere
at some point.  To verify this, note that the condition $\sin\alpha = 0$
(from eq.~(\ref{eq25}))
implies that $\alpha = \pi$ when the wheel is outside the sphere.
Then eqs. (\ref{eq27}-\ref{eq28d}) indicate, for example, that if the
wheel starts from rest at the top of the sphere it loses contact with
the sphere when
\begin{equation}
\cos\phi = {2 \over 3 + 2k}.
\label{eq82}
\end{equation}
The result for a particle sliding on a sphere ($k = 0$) is well known.

For horizontal orbits, eqs.~(\ref{eq29}-\ref{eq32}) are still valid, but
the condition that friction have an outward component is now
\begin{equation}
r'\Omega^2 < g\cot\theta,
\label{eq83}
\end{equation}
and hence
\begin{equation}
\cot\alpha < (2k + 1 + k(a/r')\cos\alpha)\cot\theta.
\label{eq84}
\end{equation}
Equation (\ref{eq29}) can be satisfied for $\alpha < \pi/2$ so long at the
radius of the wheel is small enough that 
$(2k + 1)r' + ka\cos\alpha$ is positive.  We must have $\theta < \pi/2$
to have $\alpha < \pi/2$ since $\alpha > \theta$, so horizontal orbits 
exist on the
upper hemisphere.  A particular solution is $\alpha = \pi/2$ for which
$\Omega = 0$; this is clearly unstable.

There is a class of orbits with $\theta < \pi/2$
and $\alpha$ very near $\pi + \theta$ that satisfy both eqs.~(\ref{eq29}) and
(\ref{eq84}).  These also appear to be unstable. 

The stability analysis of the preceding section holds formally for wheels
outside spheres, but the restriction there to the case of $\theta = 90^\circ$
provides no insight into the present case.

\section{Sphere Rolling Inside a Fixed Sphere}

The case of a sphere rolling on horizontal orbits inside a fixed sphere has 
been treated by Milne \cite{Milne}.  For completeness, we give an analysis
for orbits of arbitrary inclination to compare and contrast with the case of
a wheel.

Again the axis normal to the orbit is called $\hat{\bf z}'$, which makes angle
$\beta$ to the vertical $\hat{\bf z}$.  The polar angle of the orbit about 
$\hat{\bf z}'$
is $\theta$ and $\phi$ is the azimuth of the point of contact between the
two spheres.  The radius of the fixed sphere is $r$.

The diameter of the rolling sphere that passes through the point of contact
must always be normal to the fixed sphere.  That is, the ``bank'' angle of
the rolling sphere is always $\theta - \pi/2$ with respect to the plane of
the orbit.

The rolling sphere has radius $a$, mass $m$ and moment of inertia $kma^2$
about any diameter.  The angular momentum is, of course, 
\begin{equation}
{\bf L}_{\rm cm} = kma^2 \vec\omega,
\label{eq41}
\end{equation}
where $\vec\omega$ is the angular velocity of the rolling sphere.

\begin{figure}[htp]  
\postscript{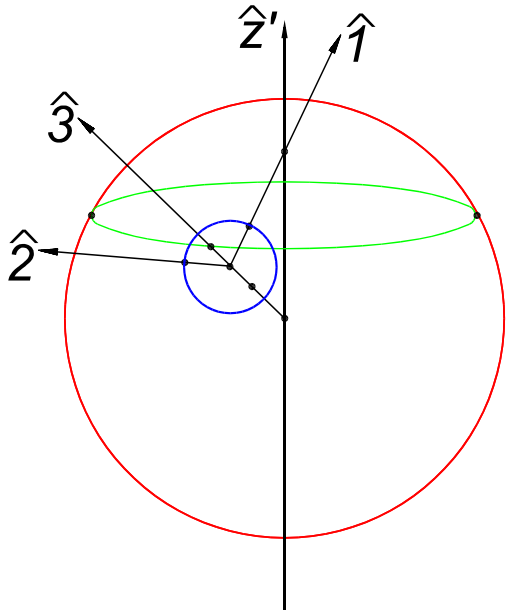}{0.4}
\begin{center}
\parbox{5.5in} 
{\caption[ Short caption for table of contents ]
{\label{fig7} Geometry illustrating the case of a sphere rolling without
slipping on a circular orbit perpendicular to the $\hat{\bf z}'$-axis inside a
 fixed sphere. The $\hat{\bf 3}$-axis is along the line of
centers of the two spheres, and passes through the point of contact.  
The $\hat{\bf 2}$-axis lies in the plane of the
orbit along the direction of motion of the center of the rolling sphere, and
axis $\hat{\bf 1} = \hat{\bf 2} \times \hat{\bf 3}$ is in the 
$\hat{\bf 3}$-$\hat{\bf z}'$ plane.
}}
\end{center}
\end{figure}

We again introduce a right-handed triad of unit vectors 
($\hat{\bf 1},\hat{\bf 2},\hat{\bf 3}$)
centered on the rolling sphere.  For consistency with the notation used
for the wheel,
axis $\hat{\bf 3}$ is directed towards the point of contact, axis $\hat{\bf 2}$ is
parallel to the plane of the orbit, and axis $\hat{\bf 1}$ is in the 
$\hat{\bf 3}$-$\hat{\bf z}'$ plane, as shown in Fig.~\ref{fig7}.  In general,
none of these vectors are body vectors for the
rolling sphere.  The center of mass of the rolling sphere lies on the
line joining the center of the fixed sphere to the point of contact, and so
\begin{equation}
{\bf r}_{\rm cm} = (r-a) \hat{\bf 3} \equiv r' \hat{\bf 3},
\label{eq42}
\end{equation}

Equations (\ref{eq2}-\ref{eq6a}) that govern the motion and describe the
rolling constraint hold for the sphere as well as the wheel.  Using
eqs.~(\ref{eq41}) and (\ref{eq42}) we can write eq.~(\ref{eq4}) as
\begin{equation}
ka{d\vec\omega \over dt} = g\hat{\bf 3} \times \hat{\bf z} 
+ r' \hat{\bf 3} \times  {d^2 \hat{\bf 3} \over dt^2}.
\label{eq43}
\end{equation}

We seek an additional expression for the angular velocity $\vec\omega$
of the rolling sphere, but 
we cannot use eq.~(\ref{eq7}) since we have not identified a body axis in
the sphere.  However, with eq.~(\ref{eq42}) the rolling constraint (\ref{eq6a})
can be written
\begin{equation}
\vec\omega = -{r' \over a} \hat{\bf 3} \times {d \hat{\bf 3} \over dt} 
+ \omega_3 \hat{\bf 3}.
\label{eq44}
\end{equation}
We can now see that $\omega_3 = \vec\omega \cdot \hat{\bf 3}$ 
is a constant by noting that $\hat{\bf 3} \cdot
d\vec\omega/dt = 0$ from eq.~(\ref{eq43}), and also $\vec\omega \cdot
d \hat{\bf 3}/dt = 0$ from eq.~(\ref{eq44}).  The freedom to chose the constant
angular velocity $\omega_3$ for a rolling sphere permits stable orbits
above the equator of the fixed sphere, just as the freedom to adjust the
bank angle $\alpha$ allows such orbits for a wheel.

Taking the derivative of eq.~(\ref{eq44}) we find
\begin{equation}
{d\vec\omega \over dt} = -{r' \over a} \hat{\bf 3} \times {d^2 \hat{\bf 3}
 \over dt^2} + \omega_3 {d \hat{\bf 3} \over dt},
\label{eq45}
\end{equation}
so the equation of motion (\ref{eq43}) can be written
\begin{equation}
(k+1)r' \hat{\bf 3} \times {d^2 \hat{\bf 3} \over dt^2} - ka \omega_3 
{d \hat{\bf 3} \over dt} = g \hat{\bf z} \times \hat{\bf 3}.
\label{eq46}
\end{equation}
Milne notes that this equation is identical to that for a symmetric top
with one point fixed \cite{Milne}, and so the usual extensive analysis of 
nutations about the stable orbits follows if desired.

We again restrict ourselves to circular orbits, for which the angular
velocity of the center of mass, and of $\hat{\bf 1}$, $\hat{\bf 2}$ and 
$\hat{\bf 3}$
is $\dot\phi \hat{\bf z}'$ where the $z'$-axis is fixed.  Then with
\begin{equation}
\hat{\bf z}' = - \sin\theta \hat{\bf 1} + \cos\theta \hat{\bf 3},
\label{eq47}
\end{equation}
we have
\begin{equation}
{d \hat{\bf 3} \over dt} = \dot\phi \hat{\bf z}' \times \hat{\bf 3} 
                   = \dot\phi \sin\theta \hat{\bf 2},
\label{eq48}
\end{equation}
\begin{equation}
{d^2 \hat{\bf 3} \over dt^2} = \dot\phi^2 \sin\theta \hat{\bf z}' \times 
   \hat{\bf 2}  + \ddot\phi \sin\theta \hat{\bf 2} 
                        =  - \dot\phi^2 \sin\theta \cos\theta \hat{\bf 1} +
\ddot\phi \sin\theta \hat{\bf 2} + \dot\phi^2 \sin^2\theta \hat{\bf 3},
\label{eq49}
\end{equation}
and hence,
\begin{equation}
\hat{\bf 3} \times {d^2 \hat{\bf 3} \over dt^2} 
= - \ddot\phi \sin\theta \hat{\bf 1} 
- \dot\phi^2 \sin\theta \cos\theta \hat{\bf 2}.
\label{eq50}
\end{equation}
With these the equation of motion (\ref{eq46}) reads
\begin{equation}
(k+1)r' \ddot\phi \sin\theta \hat{\bf 1} +
[(k+1)r' \dot\phi^2 \cos\theta + ka \omega_3 \dot\phi] \sin\theta \hat{\bf 2}
= - g \hat{\bf z} \times \hat{\bf 3}.
\label{eq51}
\end{equation}

We can use eq.~(\ref{eq22}) for $\hat{\bf z} \times \hat{\bf 3}$ if we substitute
$\alpha = \theta - \pi/2$ for the rolling sphere:
\begin{equation}
\hat{\bf z} \times \hat{\bf 3} = - \sin\beta\sin\phi \hat{\bf 1} +
    (\sin\theta\cos\beta - \cos\theta\sin\beta\cos\phi) \hat{\bf 2}.
\label{eq51a} 
\end{equation}
The components of the equation of motion are then
\begin{equation}
(k+1)r' \ddot\phi \sin\theta = \sin\beta\sin\phi,
\label{eq52}
\end{equation}
\begin{equation}
[(k+1)r' \dot\phi^2 \cos\theta + ka \omega_3 \dot\phi] \sin\theta =
g\cos\theta\sin\beta\cos\phi - g\sin\theta\cos\beta.
\label{eq53}
\end{equation}

The two equations of motion are not consistent in general.  To see this,
take the derivative of eq.~(\ref{eq53}) and substitute $\ddot\phi$ from
eq.~(\ref{eq52}):
\begin{equation}
ka\omega_3 \sin\beta \sin\phi = -3(k+1)r' \dot\phi \cos\theta \sin\beta
\sin \phi.
\label{eq54}
\end{equation}
While this is certainly true for $\beta = 0$ (horizontal orbits), for
nonzero $\beta$ we must have $\dot\phi \cos\theta$ constant since
$\omega_3$ is constant.  Equation~(\ref{eq54}) is satisfied for
$\theta = \pi/2$ (great circles), but for arbitrary $\theta$ we would
need $\dot\phi$ constant which is inconsistent with eq.~(\ref{eq52}).
Further, on a great circle eq.~(\ref{eq53}) becomes $ka\omega_3 \dot\phi = 
-g\cos\beta$.   This is inconsistent with eq.~(\ref{eq52}) unless
$\beta = \pi/2$ (vertical great circles) and $\omega_3 = 0.$

In summary, the only possible closed orbits for a sphere rolling within a
fixed sphere are horizontal circles and vertical
great circles.  

We remark further only on the horizontal orbits.  For these $\dot\phi
\equiv \Omega$ is constant according to eq.~(\ref{eq52}). Equation
(\ref{eq53}) then yields a quadratic equation for $\Omega$:
\begin{equation}
(k + 1)r'\Omega^2\cos\theta + ka\omega_3\Omega + g = 0,
\label{eq55}
\end{equation}
so that there are orbits with real values of $\Omega$ provided
\begin{equation}
(ka\omega_3)^2 \geq 4(k+1)gr'\cos\theta.
\label{eq56}
\end{equation}
This is satisfied for orbits below the equator ($\theta > \pi/2$) for
any value of the ``spin'' $\omega_3$ of the sphere (including zero),
but places a lower limit on $\abs{\omega_3}$ for orbits above the equator.  For
the orbit on the equator we must have $\Omega = -g/(ka\omega_3)$ so a
nonzero $\omega_3$ is required here as well.

The contact force ${\bf F}$ is given by
\begin{equation}
{\bf F}/m = (g + r'\Omega^2\cos\theta)\sin\theta \hat{\bf 1} -
 (r'\Omega^2\sin^2\theta - g\cos\theta) \hat{\bf 3},
\label{eq57}
\end{equation}
using eqs. (\ref{eq2}) and (\ref{eq54}).  For the rolling sphere to remain
in contact with the fixed sphere there must be a positive 
component of ${\bf F}$ pointing toward the center of the fixed  sphere.  Since 
axis $\hat{\bf 3}$ is radial outward from the fixed sphere, we require that
$F_3$ be negative, and hence
\begin{equation}
r'\Omega^2\sin^2\theta > g\cos\theta.
\label{eq58}
\end{equation}
This is always satisfied for orbits below the equator.  For orbits well
above the equator this requires a larger value of $\abs{\omega_3}$ than does
eq.~(\ref{eq56}).  To see this, suppose $\omega_3$ is exactly at the
minimum value allowed by eq.~(\ref{eq56}), which implies that 
$\Omega = -ka\omega_3/(2(k+1)r'\cos\theta)$.  Then eq.~(\ref{eq58})
requires that $\tan^2\theta > k+1$.  So for $k = 2/5$ and at angles
 $\theta < 50^\circ$
larger values of $\abs{\omega_3}$ are needed to satisfy eq.~(\ref{eq56}) than 
to satisfy eq.~(\ref{eq56}).
However, there are horizontal orbits at any $\theta > 0$ for $\abs{\omega_3}$
large enough.

\section{Sphere Rolling Outside a Fixed Sphere}

This case has also been treated by Milne \cite{Milne}.  A popular example
is spinning a basketball on one's fingertip.

Equations eq.~(\ref{eq55}) and (\ref{eq56}) hold with the substitution that
$r' = r + a$.  The condition on the contact force becomes
\begin{equation}
r'\Omega^2\sin^2\theta < g\cos\theta,
\label{eq59}
\end{equation}
which can only be satisfied for $\theta < \pi/2$.  While eq.~(\ref{eq56})
requires a large spin $\abs{\omega_3}$, if it is too large eq.~(\ref{eq59})
can no longer be satisfied in view of the relation (\ref{eq55}).  For any
case in which the orbit exists a perturbation analysis shows that the
motion is stable against small nutations \cite{Milne}.

\end{document}